\documentclass[12pt]{emulateapj}
\usepackage{graphicx}
\usepackage{times}
\usepackage{natbib}
\usepackage{amsfonts}
\usepackage{amsmath}
\usepackage{amsbsy}
\usepackage{bm}
\usepackage{url}
\usepackage{microtype}
\usepackage{rotating}
\usepackage{booktabs}
\usepackage{threeparttable}
\usepackage{tabularx}
\usepackage{subfigure}

\newcommand{\project}[1]{\textsl{#1}}
\newcommand{\fermi}{\project{Fermi}}
\newcommand{\swift}{\project{Swift}}
\newcommand{\chandra}{\project{Chandra}}
\newcommand{\rxte}{\project{RXTE}}
\newcommand{\integral}{\project{INTEGRAL}}
\newcommand{\nustar}{\project{NuSTAR}}
\newcommand{\ginga}{\project{Ginga}}

\newcommand{\likelihood}{{\mathcal L}}

\def \deg{$^\circ$}
\def \src{V404~Cyg}

\usepackage{amssymb,xcolor}
\newcommand{\SEM}[1]{\textbf{\color{blue} #1}}

\begin{document}

\title{Detection of Very Low-Frequency Quasi-Periodic Oscillations in the 2015 Outburst of V404 Cygni}

\author{D.~Huppenkothen\altaffilmark{1, 2, 3}, G.~Younes \altaffilmark{4}, A.~Ingram\altaffilmark{5}, C.~Kouveliotou\altaffilmark{4}, E.~G{\"o}{\u g}{\"u}{\c s}\altaffilmark{6}, M.~Bachetti\altaffilmark{7}, C.~S\'{a}nchez-Fern\'{a}ndez\altaffilmark{8}, J.~Chenevez\altaffilmark{9}, S.~Motta\altaffilmark{10}, M.~van der Klis\altaffilmark{5}, J.~Granot\altaffilmark{11}, N.~Gehrels\altaffilmark{12}, E.~Kuulkers\altaffilmark{8}, J.A.~Tomsick\altaffilmark{14}, D.J.~Walton\altaffilmark{15,16}}
 
  \altaffiltext{1}{Center for Data Science, New York University, 726 Broadway, 7th Floor, New York, NY 10003}
  \altaffiltext{2}{Center for Cosmology and Particle Physics, Department of Physics, New York University, 4 Washington Place, New York, NY 10003, USA}
  \altaffiltext{3}{E-mail: daniela.huppenkothen@nyu.edu}
  \altaffiltext{4}{Department of Physics, The George Washington University, Washington, DC 20052, USA}
  \altaffiltext{5}{Anton Pannekoek Institute, University of Amsterdam, Science Park 904, 1098 XH Amsterdam, The Netherlands}
  \altaffiltext{6}{Sabanc\i~University, Orhanl\i-Tuzla, \.Istanbul  34956, Turkey}
  \altaffiltext{7}{INAF/Osservatorio Astronomico di Cagliari, via della Scienza 5, I-09047 Selargius (CA), Italy}
  \altaffiltext{8}{European Space Astronomy Centre (ESA/ESAC), Science Operations Department, E-28691 Villanueva de la Ca\~nada, Madrid, Spain}
  \altaffiltext{9}{DTU Space - National Space Institute, Technical University of Denmark, Elektrovej 327-328, 2800 Lyngby, Denmark}
  \altaffiltext{10}{University of Oxford, Department of Physics, Astrophysics, Denys Wilkinson Building, Keble Road, OX1 3RH, Oxford, United Kingdom}
 \altaffiltext{11}{Department of Natural Sciences, The Open University of Israel, 1 University Road, P.O. Box 808, Ra’anana 43537, Israel}
  \altaffiltext{12}{Astrophysics Science Division, NASA Goddard Space Flight Center, Mail Code 661, Greenbelt, MD 20771, USA}
  \altaffiltext{13}{European Space Astronomy Centre (ESA/ESAC), Science Operations Department, E-28691 Villanueva de la Ca\~nada, Madrid, Spain}
  \altaffiltext{14}{Space Sciences Laboratory, 7 Gauss Way, University of California, Berkeley, CA 94720-7450, USA}
  \altaffiltext{15}{Cahill Center for Astronomy and Astrophysics, California Institute of Technology, Pasadena, CA 91125, USA}
  \altaffiltext{16}{Jet Propulsion Laboratory, California Institute of Technology, 4800 Oak Grove Drive, Pasadena, CA 91109, USA}
  
\begin{abstract}
In June 2015, the black hole X-ray binary (BHXRB) V404 Cygni went into outburst for the first time since 1989. 
Here, we present a comprehensive search for quasi-periodic oscillations (QPOs) of V404 Cygni during its recent outburst, utilizing data from six instruments on board five different X-ray missions: \swift/XRT, \fermi/GBM, \chandra/ACIS, \integral's IBIS/ISGRI and JEM-X, and \nustar. 
We report the detection of a QPO at $18\,\mathrm{mHz}$ simultaneously with both \fermi/GBM and \swift/XRT, another example of a rare but slowly growing new class of mHz-QPOs in BHXRBs linked to sources with a high orbital inclination. 
Additionally, we find a duo of QPOs in a \chandra/ACIS observation at $73\,\mathrm{mHz}$ and $1.03\,\mathrm{Hz}$, as well as a QPO at $136\,\mathrm{mHz}$ in a single \swift/XRT observation that can be interpreted as standard Type-C QPOs. 
Aside from the detected QPOs, there is significant structure in the broadband power, with a strong feature observable in the \chandra\ observations between $0.1$ and $1\,\mathrm{Hz}$. We discuss our results in the context of current models for QPO formation. 
\end{abstract}


\section{Introduction} 
\label{section:intro}

Black hole X-ray binaries (BHXRBs), systems consisting of a stellar-mass black hole and a main sequence or giant companion star, are one of the prime targets for studying accretion physics and strong gravity. These sources exhibit strong variability patterns at X-ray wavelengths, on timescales of milliseconds to weeks. The study of these variability signatures is of particular interest, because it presents a direct link to changes in the mass accretion rate as well as the geometry of the system. BHXRBs are also crucial for understanding super-massive black holes (SMBHs) at the centers of galaxies, which are $\sim 10^5-10^8$ more massive than their stellar-mass counterparts. The relevant timescales in these systems are of the order of minutes for the shortest dynamical timescales or years and longer for typical viscous timescales, and therefore too long to be directly observed with high precision \citep{mchardy2006,koerding2007,uttley2014}.

The power spectrum of a BHXRB consists of one or multiple broad-band noise components, generally well modelled by a mixture of Lorentzians with a location parameter equal to zero, or a (broken) power law \citep{belloni1990,smith1997,berger1998,nowak1999,pottschmidt2003}.

Additionally, many black hole binaries show the presence of strong quasi-periodic oscillations (QPOs): sharply pointed features in the power spectrum that can be modelled with narrow Lorentzians, but are too wide to be considered strictly coherent (periodic). 
Both broad-band noise and QPOs are believed to be stochastic processes produced by the accretion flow, although their exact physical origin is still debated.

During outbursts, usually lasting weeks to months, variability patterns change considerably following changes of the energy spectrum of the source  \citep{done2007}. Many sources generally follow a q-shaped track in the hardness-intensity diagram, which plots spectral hardness against the flux of a source over the course of an outburst \citep{belloni2005}.
While in the low hard state, the energy spectrum consists of a power law component generally dominating over the disk multi-colour blackbody component; at the same time their power density spectra usually show strong broadband noise up to $\sim 10 \,\mathrm{Hz}$. The precise origin of this power-law component in the energy spectrum is as of yet unknown, but is believed to originate in the inner regions close to the black hole, either from a hot inner flow \citep[e.g.][]{narayan1995,esin1997} or the base of an outflowing jet \citep[e.g.][]{markoff2005,miller2007}. 

As the source brightens and softens, moving through the hard intermediate state and the soft intermediate state, the spectrum becomes dominated by the multi-colour blackbody component associated with the accretion disk. At the same time, the amplitude of the broadband noise components decreases, and strong QPOs may appear at low frequencies between $0.1\,\mathrm{Hz}$ and $1\,\mathrm{Hz}$ \citep[e.g.][]{munosdarias2011,belloni2011,heil2015}. 
The frequency of these components increases as the source continues to soften, reaching as high as $\sim 10\,\mathrm{Hz}$ in the intermediate soft state, before vanishing as the source moves fully into the high soft state. 
In the high soft state, where emission is almost exclusively due to the accretion disk in the form of a multi-colour blackbody, variability in the source is strongly suppressed, leaving only weak broadband noise, before the source moves back into the hard state. 

V404 Cygni is one of the best-studied BHXRB systems in quiescence due to its proximity. It was first identified in optical observations in 1938 as Nova Cyg 1938 and subsequently misclassified as a nova \citep{wachmann1948} until \ginga\ observations during its second known outburst in 1989 \citep{makino1989} identified the X-ray source GS $2023+338$ at that position. \citet{richter1989} identified a third outburst associated with V404 Cygni in 1956 after a systematic review of photographic plates.

V404 Cygni orbits its companion in a relatively wide orbit of 6 days \citep{casares1992}. Its dynamical mass measurement of $\sim 9 \mathrm{M}_{\sun}$ \citep{casares1992,khargharia2010} quickly led to the source's identification as the most likely stellar-mass black hole candidate known. Its orbital inclination of ${67^{\circ}}^{+3}_{-1}$ places it in the group of high-inclination X-ray binaries \citep{shahbaz1994,khargharia2010}. 
With its close distance to Earth of only $2.39\pm 0.14$ kpc \citep{millerjones2009}, its very high luminosity of up to $\sim 5 \times 10^{38} \mathrm{erg}\,\mathrm{s}^{-1}$ and extreme brightness variations of a factor of $\sim 500$ on timescales of seconds \citep{kitamoto1989}, V404 Cygni is an excellent target for studying variability behaviour at high resolution. 
Variability studies with Ginga during the 1989 outburst revealed no credible QPO detection, but showed strong broad-band noise as well as features in the power spectrum that, while too broad to fall into the standard definition of a QPO, were nevertheless localized in frequency \citep{oosterbroek1997}.

The source went into outburst for a fourth time in June 2015: first detected by the \swift\ Burst-Alert Telescope (BAT, \citealt{barthelmyatel2015}), it was swiftly followed up by numerous instruments both from the ground and from space \citep[e.g.][]{kuulkers2015,golenetskii2015,burns2015, negoro2015, gazeas2015, mooley2015a, tetarenko2015, garner2015}.
The outburst lasted for almost a month, and the source returned to quiescence in early August \citep{sivakoffatel2015}. 
The outburst itself was characterized by bright flaring on timescales of hours, which could be observed both in optical \citep{gandhi2016} and in X-rays up to $\sim 400 \,\mathrm{keV}$ in \integral\ \citep{ferrigno2015,natalucci2015,rodriguez2015} and \fermi \citep{jenkegcn2015,jenke2016}. During these flares, the source reached nearly $0.4\,\mathrm{L}_{\mathrm{Edd}}$. A marginal QPO has previously been reported in a \swift/XRT observation \citep{mottaatel2015,radhika2016}. While no QPOs have been detected in the \fermi/GBM data for the duration of the entire outburst, \citet{jenke2016} report the detection of broad Lorentzian features typical of a BHXRB in the hard state.

Here, we report on the results of an X-ray timing study of V404 Cygni throughout its outburst in June/July 2015. 
We utilize data from the X-ray Telescope (XRT) onboard \swift\ as well as the \fermi\ Gamma-Ray Burst Monitor (GBM) to study the timing evolution of the source and search for the presence of QPOs. 
Additionally, we include two long, public \chandra\ Advanced CCD Imaging Spectrometer (ACIS) observations  to increase our coverage of the source, as well as several observations taken with the JEM-X and IBIS/ISGRI instruments onboard \integral and one observation taken with \nustar (see Table \ref{tab:obsoverview} for details).\begin{table}[hbtp]
\renewcommand{\arraystretch}{1.3}
\footnotesize
\caption{Overview of the observations considered in this work}
\begin{threeparttable} 
\begin{tabularx}{9cm}{p{2.5cm}p{1.5cm}p{2.0cm}p{1.5cm}}
\toprule
\bf{Instrument} & \bf{Number of observations} & \bf{Number of uninterrupted light curves} & \bf{Total observation time [ks]} \\ \midrule
\swift/XRT & $35$ & $49$ & $46.1$ \\
\fermi/GBM & $34$ & $34$ & $28.7$\\
\chandra/ACIS & $2$ & $14$  & $47.8$\\
\nustar & $1$ & $1$ & $64.4$ \\
\integral/JEM-X & $53$ & $52$ &$168.1$ \\
\integral/ISGRI &  $53$ & $52$ &  $168.1$ \\
\bottomrule
\end{tabularx}
\end{threeparttable}
\label{tab:obsoverview}
\end{table} 
We find four distinct QPOs between $18\,\mathrm{mHz}$ and $1.03 \,\mathrm{Hz}$ in three of the data sets considered in this work. 

We also track the evolution of the broadband noise components in the \integral/JEM-X data as well as the \chandra\ observations, and find that the power spectrum is well-modelled by three Lorentzian components in both instruments, but the centroid frequency of these components increases between the \integral\ and the \chandra\ observations.
We discuss the likely origin of the variability patterns we see as well as implications for future studies. 

\section{Observations}
\label{section:obs}

\subsection{\swift}

\swift/XRT \citep{burrows2005} observed \src\ throughout its 2015 outburst  in $35$ separate pointings (see Table~\ref{tab:obsoverview}). 
For temporal analyses purposes, we only focus on the XRT data taken in Windowed Timing (WT) mode. 
The configuration of this mode of observation was w2, achieving a temporal resolution of 1.78~ms. 
The individual \swift/XRT observations lasted between $0.71$ and $5.66$ ks, split into multiple good time intervals (GTIs) of $\sim0.1-4$~ks \citep[see also ][]{radhika2016}. 
For much of the time that \swift/XRT observed, the source had a count rate significantly above $150\,\mathrm{counts}\,\mathrm{s}^{-1}$, thus pile-up significantly affects the data, even when excluding the central pixels \citep[see ][]{kalamkar2013}.
Pile-up changes the shape of the spectrum by registering two simultaneously incident low-energy photon as a single higher-energy photon. 
It also affects the count rates: the brighter the source, the higher the fraction of photons lost. 
Thus, pile-up effects may weaken the observed variability. 

In practice, there are three consequences to analysing piled-up data: (1) the powers in the periodogram are not strictly Poissonian. However, that effect is only important at much higher frequencies than the ones considered in this study; (2) the suppression of variability makes it more difficult to detect a QPO at high count rates; (3) the quoted fractional rms amplitudes should be considered with caution, and perhaps rather be regarded as lower limits. 

We processed the raw data using standard procedures, bary-center corrected the resulting light curves and filtered for the energy range 2-10~keV. The log of the \swift/XRT observations is summarized in Table~\ref{tab:obsoverview}.

We chose all segments with a duration of at least $256\,\mathrm{s}$. 
The duration was chosen as a trade-off between excluding too much data in shorter segments, and having segments long enough that we can probe timescales down to $10 \,\mathrm{mHz}$. 
The total duration spent on the source, corrected for GTIs, is $47.59 \,\mathrm{ks}$ in $49$ uninterrupted segments. 
Applying our constraint of having only segments with at least $256\,\mathrm{s}$ duration excludes $9$ of these segments with a total duration of $1526$ seconds, leaving us with $40$ usable segments of $46.065\,\mathrm{ks}$ duration. An overview of the light curves is presented in Figure \ref{fig:data_overview}, uppermost panel.

\subsection{\chandra}
Although \chandra\ observed \src\ on several occasions during and in the aftermath of the outburst, here we only focus on the two observations that were taken in continuous clocking mode (CC-mode), using High Energy Transmission Grating (HETG, \citealt{canizares2005}). 
The observations with IDs $17696$ and $17697$ were taken on 22 June 2015, 13:39:21 UT and 23 June 2015, 21:25:32 UT, for a total of 21 and 25~ks, respectively \citep[see also ][]{king2015}.

The HETG comprises two sets of gratings, the medium energy grating (MEG), operating in the energy range of $0.4-7$~keV, and the high energy grating (HEG) with energy coverage in the range of $0.8-10$~keV. 
Each grating spectrum is dispersed along the ACIS-S CCDs into positive and negative spectral orders. 
For these observations, only the HEG$-$ and MEG$+$ orders are recorded. The CC-mode collapses the usual 2-D image into 1-D, resulting in much improved temporal resolution of 2.85~ms.

A few flaring episodes are observed during both observations, some of which reach extremely high count rates (of $\sim 800\,\mathrm{counts}\,\mathrm{s}^{-1}$). 
We estimated the pile-up fraction of the source by calculating the count rate landing in a $3\times3$ pixel island and the CCD readout frame time. 
We find that pile-up is not an issue during these observations with a maximum pile-up fraction of $\sim15\%$ at the peak, lasting for a few seconds of the strongest flare.

We checked the background level in each of the observations from the order sorting plots, which display the energies of the dispersed events versus the ratios of these energies over the event positions on the grating arm \citep[see e.g.\ ][]{younes2015}. 
Point source photons should distribute tightly around the extraction order, while background photons scatter around. We find that the background is energy dependent and strongest at the edges of the energy coverage of the arms. 
Hence, for both observations, we custom filter for energy, only using events in the energy range $2-10$ keV for HEG as well as the order sorting ratio to maximize the S/N ratio in the data. Both observations were barycenter-corrected. 

To use the grating arms in our timing analysis, the photon-assigned times needed to be corrected for their diffraction angle, which is directly proportional to the grating time offset with respect to the zeroth order \citep{younes2015}.  
This time offset, $\delta t$, relative to the zeroth order location is

\begin{equation}
\delta t=-\frac{\sin(tg\_r_{\rm i})\times X_{\rm R}\times\sin\alpha_{\rm i}}{\Delta_{\rm p}}~t_{\rm p}~[{\rm s}]
\end{equation}

\noindent where $tg\_r_{\rm i}$ is the diffraction angle of each photon $i$, $X_{\rm R}$ is the Rowland spacing, $\alpha_{\rm i}$ is the grating clocking angle, $\Delta_{\rm p}$ is the pixel size, and $t_{\rm p}$ is the read time per row (2.85~ms).

The source's brightness required placement of the zeroth grating order off the CCD in order to protect the detector \citep[see][for more details]{king2015}. At the same time, dithering produces a prominent quasi-periodic oscillation (visible both in the light curves as well as the periodograms at $746\,\mathrm{s}$. Dithering mostly affects counts near the source (i.e. closer to the edges of the detector) and at higher energies. To mitigate the effect, we exclude the MEG entirely from the analysis. This reduces the number of photons in the observations by a factor of two, but removes the spurious QPO due to dithering from both observations.

\begin{figure*}[htbp]
\begin{center}
\includegraphics[width=16cm]{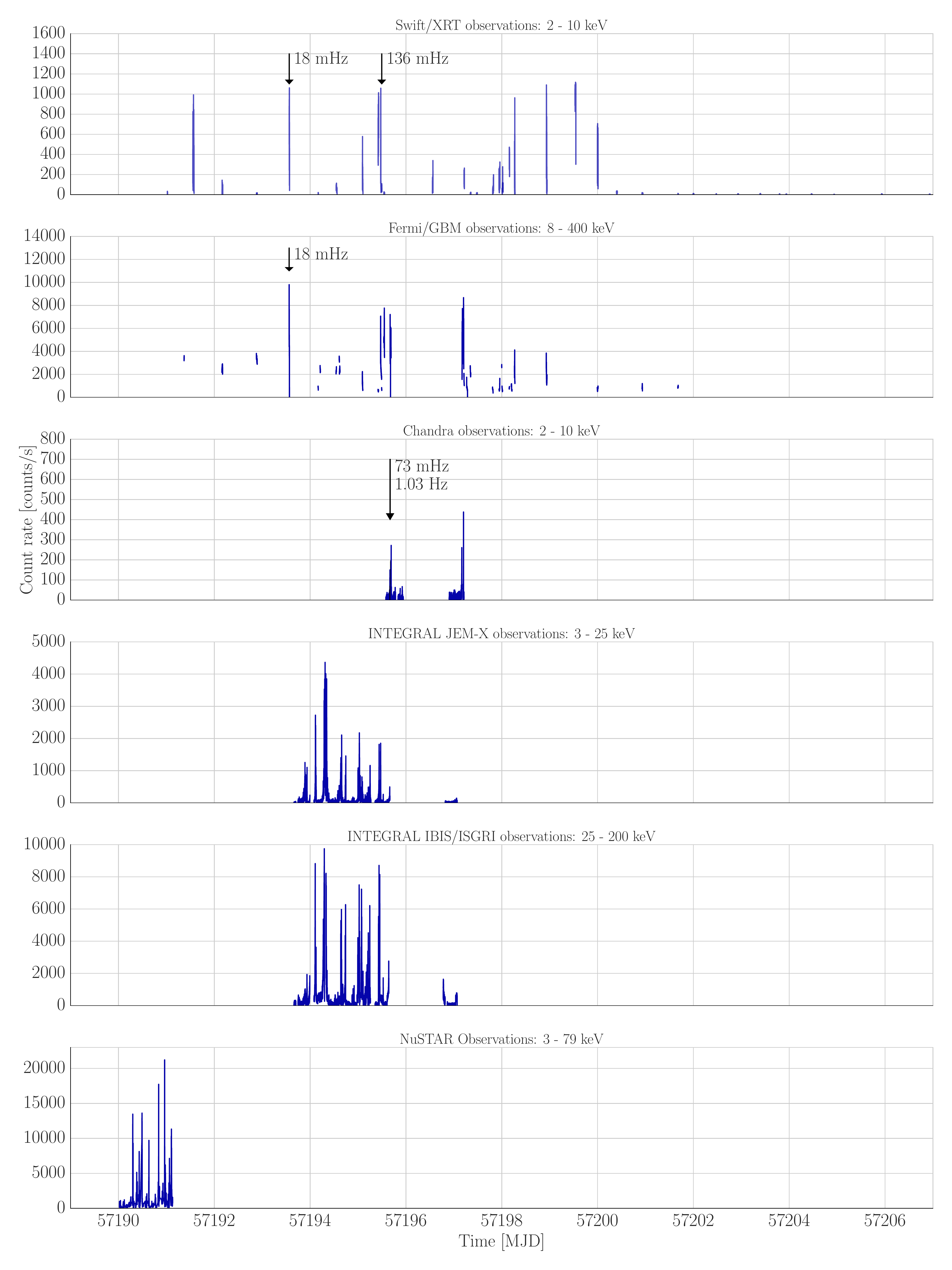}
\caption{An overview of the observations used in this paper. The data plotted are the actual light curves for \swift/XRT (first panel), \fermi/GBM (second panel), \chandra/ACIS (third panel), \integral/JEM-X and IBIS-ISGRI (fourth and fifth panel), and \nustar\ (bottom panel). The figure demonstrates the different cadences and observing durations for the different instruments used. \swift/XRT (first panel) observed the source in somewhat regular intervals, but for very short observations, making the light curves look compressed in this figure. \chandra\ (third panel) and \nustar\ observed in long, but rare pointings. \integral\ (fourth and fifth panel) observed almost continuously, though here we picked those orbits here that are close to the high-time resolution \chandra\ pointings to be able to compare the two instruments. While \fermi/GBM could, in principle, observe most of the time, the motion of the spacecraft and its non-pointed observations make timing difficult for long stretches, thus we have restricted ourselves here to triggers and intervals coincident with observations taken with the other spacecraft.
Clearly visible in the top two panels are the large flux variations within and in between pointings, sometimes by more than a factor of $100$. The observations where QPOs are found are indicated with arrows, with the respective QPO frequencies stated as well. For detailed light curves of these observations, please see Figures \ref{fig:lowfreq_qpo_lcs} (for the 18 mHz signal in \swift/XRT and \fermi/GBM), \ref{fig:chandra_qpo_lcs} (for the \chandra\ QPOs) and \ref{fig:xrt_qpo2} for the second potential QPO observed in \swift/XRT.}
\label{fig:data_overview}
\end{center}
\end{figure*}

The \chandra\ observations are summarized in Table~\ref{tab:obsoverview} and presented in Figure \ref{fig:data_overview}, third panel.
Because the \chandra\ GTIs cut off when the source is too bright, many of the continuous light curves either begin or end with a sharply 
rising tail. 

In order to avoid windowing effects introduced by these trends in the data, we visually inspected all \chandra/ACIS light curves and clipped them to remove any sharply rising trends at the beginning or end of each light curve.

\subsection{\fermi}

The \fermi/Gamma-ray Burst Monitor (GBM, \citealt{meegan2009}) has a continuous broadband energy coverage (8 keV-40 MeV) of the Earth un-occulted sky. 
It consists of 12 NaI detectors (8-1000 keV) and 2 Bismuth Germanate (BGO) detectors (0.2-40 MeV). \fermi/GBM continuously records photon arrival times in the 
form of time-tagged event (TTE) data as part of the \textit{daily data} products. The same TTE data are also published in smaller, more manageable files, optimized for the detection and analysis of Gamma-ray Bursts (GRBs) whenever the instrument triggers on a bright source, including photon arrival times from $30\,\mathrm{s}$ before the trigger to $600\,\mathrm{s}$ after the trigger.
\fermi/GBM triggered on a large number of the \src\ flares during the outburst \citep{jenke2016}, while many others are seen in the daily data, which did not trigger either because of low flux and/or fluence that did not reach the triggering threshold or the fact that they took place in the 600~s trigger-free window after each trigger.
Only three of these triggers coincide with \swift/XRT observations: triggers $150620567$,  $150619165$ and $150619173$ are simultaneous with \swift/XRT observations $00031403040$ and $00031403042$. 
However, because of \fermi's continuous observing mode, there is high-resolution continuous time-tagged event (CTTE) data with a time resolution of $2~\mu\mathrm{s}$ available for all but four \swift/XRT pointings. 

For this study, we focus on the GBM NaI data during time intervals that are simultaneous with the \swift/XRT data introduced above. 
For that purpose, we first check the angles of the 12 NaI detectors during each of these intervals to establish which combination of detectors to use according to their angle to the source and blockage status (blockage due to the spacecraft, radiation panels, and/or the Large Area Telescope instrument). 
Most of these time intervals had durations long enough to be affected by the \fermi\ spacecraft movement in the sky, which altered the detector angles to the source. 
Hence, we follow the detector angles throughout the time intervals considered in a time bin-size of 300~s, which is usually short enough to avoid major changes to the detector angles to the source. 
During these 300~s long intervals, we only consider the detectors with angles $<50$\deg. 

There are two \fermi/GBM triggers coincident with \chandra\ observation $17696$: triggers $150622672$ and $150622684$.
While we could have extracted continuous \fermi/GBM data for the intervals simultaneous with the \chandra\ observations as well, this is technically much more challenging than for the \swift\ data. 
\fermi's constant sweeping motion across the sky means that the background in the detectors changes continuously on time scales of hundreds of seconds. 
Additionally, every $\sim1000\,\mathrm{s}$ or so, the source vanishes from the field of view of one detector and appears in the field 
of view of another. 
This leads to data segments that vary both in background and in sensitivity. 
The timing properties of this type of data are not very well understood and likely require more complex methodology in order to be taken into account properly. 
The duration of the \swift/XRT light curves is generally short enough to be covered with the same detectors in \fermi/GBM; this is not true for the \chandra\ data. 
Therefore, while we could extract CTTE data simultaneous with the \swift/XRT data, we chose to restrict ourselves to \fermi/GBM triggers simultaneous with the \chandra\ observations instead of the full CTTE data set covering the entire \chandra\ observation interval. At the same time, these triggers represent the brightest and likely most interesting intervals when the source was active.

All photons extracted are in the $8-400 \,\mathrm{keV}$ range and are barycenter-corrected to the center of mass of the solar system. 
The lower edge of the energy range used is determined by the detector \citep{meegan2009}; the upper edge is a trade-off between including as much source emission as possible and avoiding the inclusion of too much background, based on the source energy spectrum \citep{jenke2016}. The data are shown in Figure \ref{fig:data_overview}, second panel from the top.

\subsection{\integral}

We analyzed a subset of the available \integral\ \citep{winkler2003} data on V404 Cyg close in time to the long \chandra\ observations in an effort to test whether any potential QPOs might be persistent over the course of the outburst. The data were collected by two of the \integral\ instruments: the \integral\ Soft Gamma-Ray Imager (ISGRI, part of IBIS \citep{ubertini2003} sensitive from $\sim$15\,keV to 1\,MeV with a total effective area of about 2600\,cm$^{2}$ \citep{lebrun2003}, and the Joint European X-ray Monitor (JEM-X, \citet{lund2003}). 
IBIS has a wide field of view  (FOV, $9^{\rm o}\times9^{\rm o}$ fully coded and $29^{\rm o}\times29^{\rm o}$ partially coded; full-width at zero response, FWZR).
JEM-X has a circular field of view with a diameter of about $13^{\rm o}$ (FWZR). 
This instrument consists of two units, which operate simultaneously. 
They are sensitive in the 3--35\,keV energy range and each detector has an effective area of about 500\,cm$^2$.\

The dataset in both JEM-X and IBIS/ISGRI consists of 46 science windows (ScWs -- corresponding to stable pointings of the satellite) during \integral\ revolution 1555, and the first seven ScWs of revolution 1556; each ScW lasts about one hour (see Table \ref{tab:obsoverview} and also \citealt{roques2015,natalucci2015,rodriguez2015}). 
Thanks to the adopted hexagonal dithering of the satellite during the whole observation the source always remained inside the 4.8 degree fully-illuminated FoV of JEM-X. 
The light curves were individually obtained for every ScW and subsequently merged together to only three light-curves, respectively covering the following time intervals (all times in UTC): 1) from 15:53 to 22:52 on June 20, 2) from 22:54 to 15:28 on June 22, and 3) from 18:39 on June 23 to 01:38 on June 24.

The \integral\ data were reduced with the {\it Off-line Scientific Analysis} software (OSA) distributed by the \integral\ Science Data Center (ISDC; \citealt{courvoisier2003}) version 10.1 released on October 4, 2014, using the OSA default parameters. 
The routines employed to analyze the ISGRI data are described in Goldwurm et al.\ (2003). 
We processed the IBIS/ISGRI data from the correction step {\sl COR} to the {\sl SPE} level, and then applied the lightcurve extraction tool  {\sl ii\_light} recommended for the extraction of light curves of bright sources, or sources that require a time resolution up to 0.1 sec.
The IBIS/ISGRI light curves were generated in the 25--200 keV band, with a time resolution of 0.1 and 1 second.

We used \integral/JEM-X to produce 3-25 keV light-curves of V404 at a resolution of 0.1s (see Figure \ref{fig:data_overview}, fourth and fifth panel from the top for a plot of the light curves from both instruments). 
The consolidated data for the JEM-X1 unit were processed with the instrument-specific analysis pipeline from the correction step to the light-curve extraction step, including the latest available calibration of the instrument. 
The source light-curves are background-subtracted and obtained from events selected at the source position, accounting for dead time and vignetting (off-axis angle in the instrument FoV) effects.

\subsection{\nustar}

We processed the \nustar\ observations (see also \citealt{walton2016}) described in Table \ref{tab:obsoverview} (ObsIDs 90102007002 and 90102007003) with the standard {\it nupipeline}\footnote{http://heasarc.gsfc.nasa.gov/docs/nustar/analysis/} shipped with HEASOFT 6.17.
We then referred the photon arrival times to the Solar System Barycenter using {\tt barycorr}. 
\nustar \citep{nustar13} has two telescopes, focusing hard X-rays (3--79 keV) to two identical focal plane modules (FPMA and FPMB) housing CZT pixel detectors.
The timing analysis was performed with MaLTPyNT (\citealt{MaltPyNT}; details below). 
We selected data only from good time intervals (GTI), with a safe interval from the start and the end of the GTI of 200 seconds that was discarded in order to avoid the effects of the increased radiation that is known to appear at the borders of GTIs. The resulting light curve is displayed in the bottom panel of Figure \ref{fig:data_overview}.
Each GTI was then split in 256-sec chunks, starting from $t_{\rm start, GTI} + 200$.

\section{Analysis Methods}
\label{section:methods}

We produced power density spectra in the rms normalization \citep{BelloniHasinger90,vanderklis1997} for all available light curves for the \swift/XRT, \fermi/GBM and \chandra/ACIS data sets in the $0.005 - 10 \,\mathrm{Hz}$ frequency range, where low-frequency QPOs are most commonly observed. For \integral/JEM-X and IBIS/ISGRI, we used a frequency range of $0.005 - 5 \,\mathrm{Hz}$, owing to instrumental restrictions on the time resolution. Searching for higher frequencies is theoretically possible, but not all instruments involved in this study allow searches in the regime where higher-frequency QPOs are typically seen. Additionally, the higher frequencies will be disproportionally affected by instrumental effects such as pile-up. Thus, in order to compare results from different instruments in a physically meaningful way, we choose here to focus on the range where low-frequency QPOs are generally seen.

In order to increase the signal-to-noise ratio, we split each light curve into shorter segments and averaged the periodograms of these segments.  
We used varying segment sizes between $16\,\mathrm{s}$ and $256\,\mathrm{s}$; the short segments allow for a high sensitivity on short timescales (high frequencies), while using long segments allows us to probe down to the lowest timescales typically observed in BHXRBs. We produced averaged periodograms per observation to track changes over the course of the outburst, as well as averaged periodograms of all light curves available during the entire outburst for a given instrument.

For the \nustar\ observation, we used the technique described by \cite{Bachetti+15} to calculate the cospectrum, a proxy of the power density spectrum, to account for the well-known effects of the \nustar\ instrumental dead time. We obtained light curves with a bin time of 0.025s separately for the focal plane modules A and B, calculated the Fourier Transform on each detector, produced the cross power density spectrum and took the real part of it, normalized following \citet{BelloniHasinger90}. 

\begin{table*}[hbtp]
\renewcommand{\arraystretch}{1.3}
\footnotesize
\caption{Overview of the priors used in the Bayesian models}
\begin{threeparttable} 
\begin{tabularx}{18cm}{p{6cm}p{6cm}p{6cm}}
\toprule
\bf{Parameter} & \bf{Meaning} & \bf{Probability Distribution} \\ \midrule
{\it Broken Power Law Broadband Noise Model}  \\ \midrule 
$\alpha_1$	& low-frequency power law index &   $\mathrm{Uniform}(0,5)$ \\
$\alpha_2$ 	& high-frequency power law index  & $\mathrm{Uniform}(0,5)$ \\
$\nu_{\mathrm{break}}$	& break frequency &  $\mathrm{Uniform}(\nu_{\mathrm{min}},\nu_{\mathrm{max}})$\tnote{\emph{a}} \\
$\log{A_\mathrm{PL}}$			& power law amplitude & $\mathrm{Uniform}(-8,8)$ \\\midrule 
{\it Lorentzian QPO Model} \\ \midrule
$\nu_0$ & centroid frequency & $\mathrm{Uniform}(\nu_{\mathrm{min}},\nu_{\mathrm{max}})$ \\
$\log{\gamma}$ & width (half-width at half-maximum, HWHM) &  $\mathrm{Uniform}(\log{\Delta\nu},\log{100\nu_0})$\tnote{\emph{b}} \\
$\log{A_{\mathrm{Lor}}}$ & amplitude of the Lorentzian & $\mathrm{Uniform}(-8,8)$ \\ \midrule
$\log{A_{\mathrm{noise}}}$ & Poisson noise amplitude &  $\mathrm{Uniform}(-8,8)$ \\
\bottomrule
\end{tabularx}
   \begin{tablenotes}
      \item{An overview over the model parameters with their respective prior probability distributions.}
      \item[\emph{a}]{$\nu_\mathrm{min}$, $\nu_\mathrm{max}$: smallest and largest frequency in the power spectrum, respectively}
     \item[\emph{b}]{$\Delta\nu$: frequency resolution; $\nu_0$: Lorentzian centroid frequency}
\end{tablenotes}

\end{threeparttable}
\label{tab:priors}
\end{table*}

This procedure is done automatically by MaLTPyNT. Since the frequencies involved are well below 100 Hz, we then corrected the measured rms by dividing it by the ratio between incident and detected photons as described in \cite{Bachetti+15}. 

Each average periodogram was modelled with a broken power law by finding the maximum-a-posteriori (MAP; the mode of the posterior distribution), employing a $\chi^2$-likelihood with varying degrees of freedom depending on the number of power spectra averaged \citep{barret2012}. For an overview of the priors on the parameters for all models employed in this work, see Table \ref{tab:priors}.
We did a first crude QPO search by dividing out the MAP model and searching for outliers in the residuals using the correct statistical distribution describing the p-value of measuring the observed outlier if no signal is present in the data \citep{groth1975}. 
The significances obtained this way are not reliable, since they generally overestimate the significance for the presence of a QPO and do not take into account the uncertainty in the broad-band noise model \citep{vaughan2010,huppenkothen2013}. 
We choose all outliers with a single-trial significance of at least $10^{-5}$ in this sample, and then used more refined, but computationally expensive methods to correctly assess significance of the candidate QPO signals. 

For a more precise assessment of the significance, we use the Bayesian QPO detection method laid out in \citet{huppenkothen2013} with a broken power law model to represent the broadband variability. 
The latter is a good estimate for all but the \chandra\ observations, which show more complex structure, largely due to the longer duration and therefore higher signal-to-noise in the individual powers in the averaged periodogram (see further below in this section for more details on the \chandra\ analysis).
 
Using Markov Chain Monte Carlo simulations (obtained with {\it emcee}, \citealt{foremanmackey2013}) of the parameters for the broadband noise model, we simulate $10000$ periodograms without a QPO, produce MAP fits for each, and compare the highest outlier in the residuals for each simulation to that of the observed periodogram.
 
This allows us to build an empirical distribution for the highest outlier and assess the significance of that outlier under the assumption that no QPO is present in the data. 
We correct all p-values for the number of periodograms searched (this includes the number of observations as well as the number of segment sizes covered; the number of frequencies is automatically taken into account by considering the highest outlier in each periodogram derived from data as well as each simulation).
For significant QPO detections, we compute the fractional rms amplitude contained in the signal, either by integrating the periodogram over the relevant frequency range, or by integrating the MAP Lorentzian component over the entire spectrum, if the signal-to-noise ratio is sufficient.

\begin{figure*}[htbp]
\begin{center}
\includegraphics[width=\textwidth]{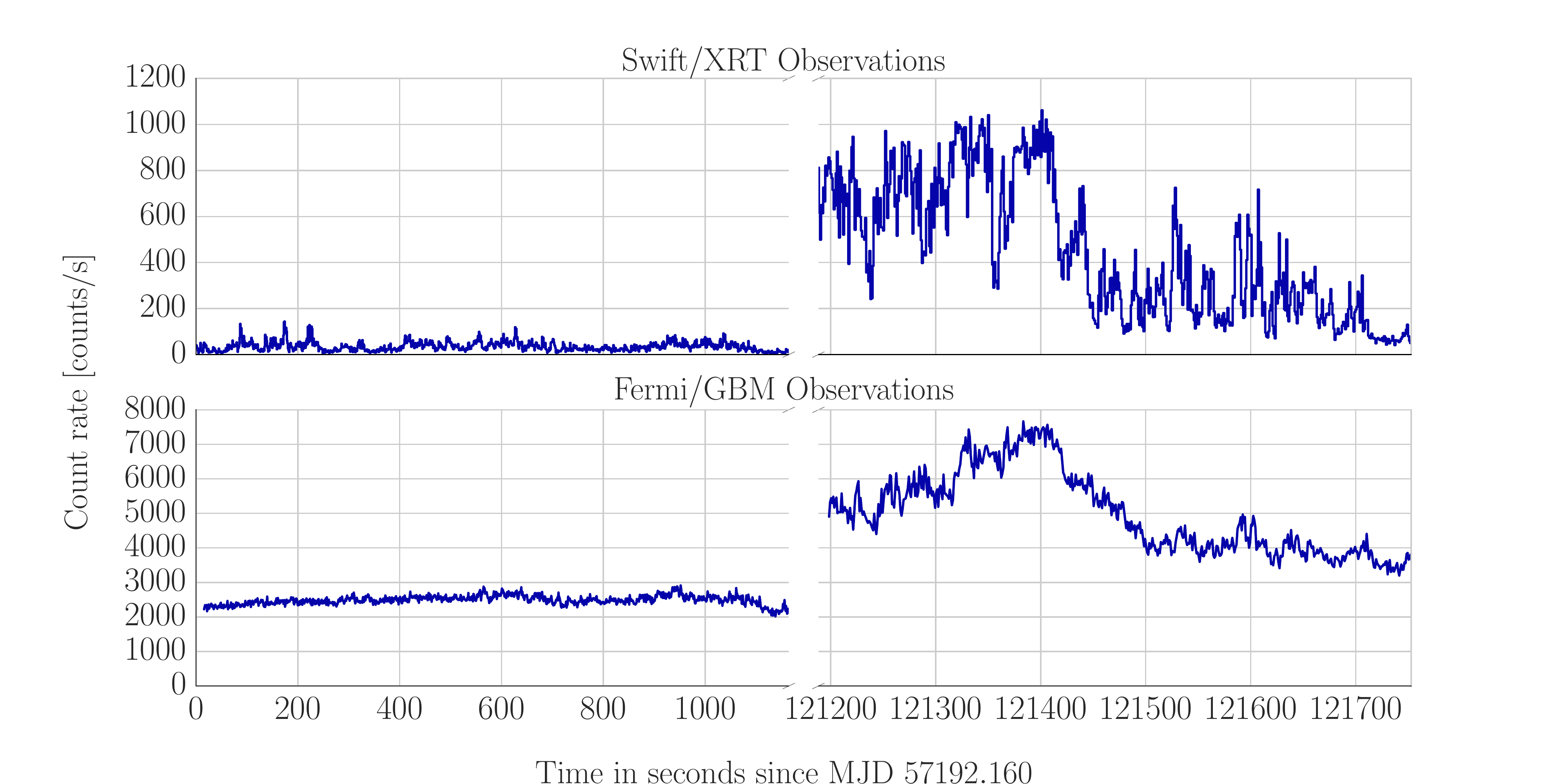}
\caption{Light curves of the two \swift/XRT triggers and simultaneous \fermi/GBM observations where the low-frequency QPO at $18\,\mathrm{mHz}$ was detected. For the corresponding power spectra, see Figure \ref{fig:lowfreq_qpo}.}
\label{fig:lowfreq_qpo_lcs}
\end{center}
\end{figure*}

For the higher-quality Chandra data, we employ a mixture of Lorentzian components to model the periodogram. 
Here, we use the Bayesian Information Criterion (BIC, \citealt{schwarz1978}; for a discussion of information criteria in an astronomical context also see \citealt{liddle2007}) to choose between competing models with different numbers of Lorentzian components making up the spectrum. The BIC is an approximation to the Bayesian evidence (also known as the marginal likelihood), i.e. the integral of the likelihood times the prior distribution over the parameter space. 
The Bayesian evidence rewards both data fit and model predictiveness, but is difficult and expensive to compute in practice. The BIC asymptotically approaches the Bayesian evidence under the conditions that the sampling distribution belongs to the exponential family, the data points are independent and identically distributed, and the number of data points is much larger than the number of parameters in the model. Formally, the BIC is defined as 

\[
\mathrm{BIC} = -2 \log{(\likelihood_\mathrm{max})} + k\log{(N)} \, ,
\]

\noindent where $\mathcal{L_\mathrm{max}}$ is the maximum likelihood estimate, $k$ is the number of parameters in the model, and 
$N$ the number of data points. 
Note that the BIC includes a correction term involving the number of parameters to prevent overfitting with complex models, thus a smaller BIC may indicate a better model fit, a lower number of parameters, or both. 
In a model comparison context, one may compare the BIC for competing models, and consider $\Delta\mathrm{BIC} > 6$ as strong evidence for the model with the smaller BIC. 
A very similar approach was recently successfully employed to detect quasi-periodic oscillations in hard X-ray data of solar flares \citep{inglis2015}. 

For all observations -- except \integral\ (see below) -- where no signal is found, we compute the sensitivity limits on the fractional rms amplitude based on the MCMC simulations.

Because of the strong aperiodic variability in the source affecting the low frequencies, these sensitivity limits will depend not only on source brightness, but also on frequency: at lower frequencies, we are more likely to miss a fairly strong QPO due to the abundance of aperiodic variability potentially masking the signal, which is less likely to occur at high frequencies, where Poisson noise becomes the dominant limiting factor.

\section{Results}
\label{section:results}

\subsection{QPO Searches}
\label{section:qposearch}

We searched for QPOs in the $0.005-10 \,\mathrm{Hz}$ frequency range in all data sets obtained with \swift/XRT, \fermi/GBM, \chandra/ACIS and \nustar\ as described in Section \ref{section:obs}. 
For the \integral/JEM-X and IBIS/ISGRI data, we searched the $0.005-5\,\mathrm{Hz}$ frequency range owing to a lower time resolution of the light curves. 
The noise properties of the coded-mask data taken with the \integral\ instruments is not well understood, but power spectra derived from \integral\ light curves are known to not exactly follow the standard $\chi^2$-distribution with $2$ degrees of freedom \citep{fuerst2010,grinberg2011}. 
This makes the search for QPOs technically more challenging than for the other instruments used in this study.
We fit the noise powers in the $1-5\,\mathrm{Hz}$ range, which are likely dominated by Poisson noise, for each light curve independently with a non-central $\chi^2_2$ distribution, leaving the non-centrality parameter free, and subsequently compare the observed powers to a non-central $\chi^2_2$ distribution using a standard two-sided Kolmogorov-Smirnov test. 
We find good agreement between the data and the test distribution for all data segments, however, the non-centrality parameter varies non-linearly with average count rate in each light curve.
We conclude that the analysis sketched out in Section \ref{section:methods} above should perform reasonably well, since the Poisson noise level at high frequencies (equivalent to the non-centrality parameter of the $\chi^2$ distribution) is a free parameter of the model. Thus we proceed with the QPO search in both \integral data sets in the same way as for the \swift/XRT and \fermi/GBM data.
However, our lack of understanding of the statistical properties of the \integral\ data leads us to be more restrictive in claiming detections from this data set (we will only accept detections in conjunction with detections at the same frequency in one of the other data sets), and no upper limits to the fractional rms amplitudes will be computed for the \integral\ observations.

\begin{figure*}[htbp]
\begin{center}
\includegraphics[width=\textwidth]{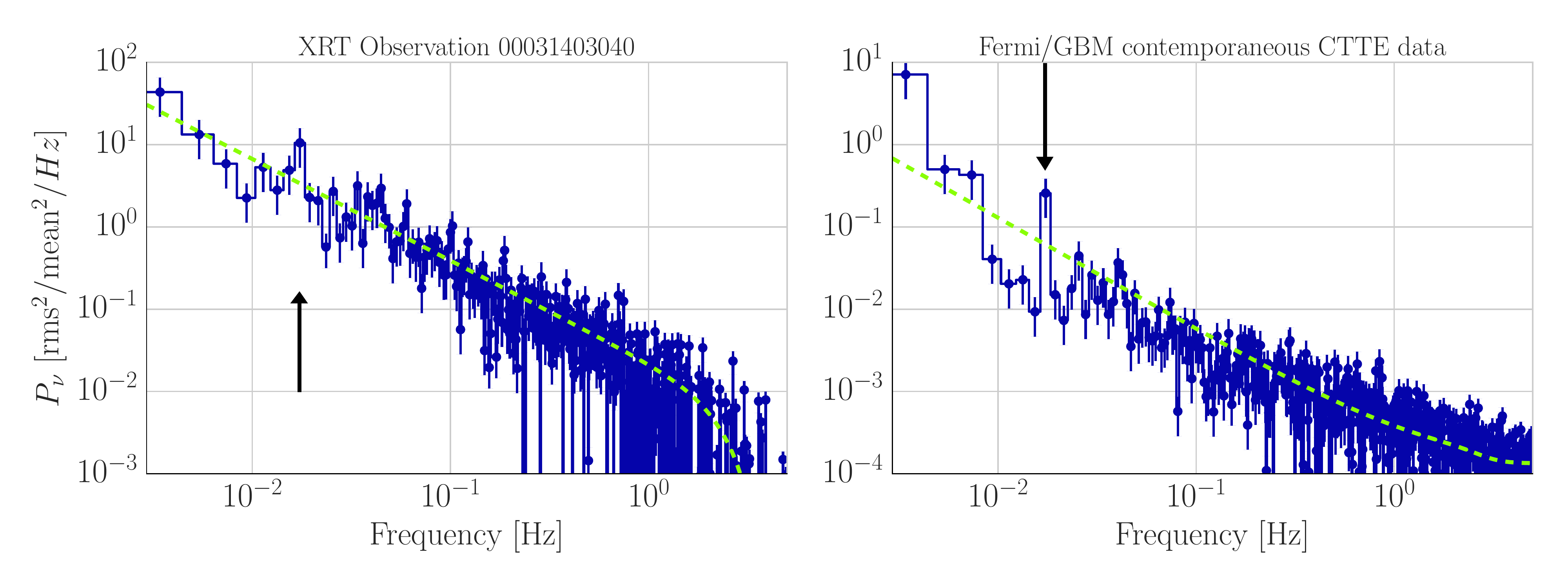}
\caption{Averaged periodograms (blue data points) for the two light curves in \swift/XRT observation $00031403040$ (left panel) and simultaneous \fermi/GBM data (right panel) showing a QPO at $18\,\mathrm{mHz}$ (black vertical lines). The green dashed line represents the MAP model of the broadband variability. The corresponding light curves are presented in Figure \ref{fig:lowfreq_qpo_lcs}.}
\label{fig:lowfreq_qpo}
\end{center}
\end{figure*}

An overview of the observations can be found in Figure \ref{fig:data_overview}. We paid special attention to time intervals where data from more than one instrument were available, since a signal detected simultaneously in two independent data sets drastically increases our confidence in its presence. We also directed close attention to those observations coincident with \fermi/GBM {\it triggers} (rather than 
untriggered CTTE data), since these present some of the brightest intervals, where the \fermi/GBM data are dominated by the source and thus less heavily affected by the background. Additionally, the improvement in photon statistics due to the brightness of the source allows us to perform more sensitive searches. 

We report the detection of a QPO in \swift/XRT observation $00031403040$ (see Figure \ref{fig:lowfreq_qpo_lcs} for a light curve) in the averaged periodogram of 6 $256\,\mathrm{s}$-long segments at $18 \,\mathrm{mHz}$ with a single-trial (classical) significance of $7 \times 10^{-5}$ and a posterior predictive p-value derived from simulations with a lower significance of $0.02\pm 0.001$.

The QPO is fairly narrow: it is confined within one frequency bin of width $\Delta\nu = 3.9 \,\mathrm{mHz}$, corresponding to a lower limit on the $q$-factor, $q = \frac{\nu_0}{\Delta\nu}$, of $\sim 4.5$ (see Figure \ref{fig:lowfreq_qpo}, left panel). 

While by itself, this candidate would not be considered convincing, we also report the detection of a similarly strong QPO 
in the simultaneous \fermi/GBM data (the light curve is also presented in Figure \ref{fig:lowfreq_qpo_lcs}) at the same frequency with a single-trial (classical) p-value of $5.68\times 10^{-5}$ and a more accurate posterior predictive p-value from simulations of $0.021\pm 0.001$ (Figure \ref{fig:lowfreq_qpo}, right panel). As with the \swift/XRT detection, the QPO in \fermi/GBM is confined within one bin for an equivalent width and $q$-factor. 

For \fermi/GBM, the broadband model appears to provide a bad fit at low frequencies, exactly where the QPO is found. In order to test whether this is the case, and whether our QPO detection is robust to changes in the broadband model, we considered alternative models for the power spectrum: similarly to our approach to the \chandra\ data, we modeled the broadband power spectrum with a superposition of two or three Lorentzians, allowing for a greater variety in power spectral shapes.  However, we find strong evidence for the simpler model and against a more complex model composed of several Lorentzians (model with two Lorentzians: $\Delta\mathrm{BIC} = 5105.53$; model with three Lorentzians: $\Delta\mathrm{BIC} = 12776.31$). This result is independent of different choices in starting parameters, thus unlikely to be due to local minima. Choosing any of the more complex models does not change the detection p-value of the QPO in \fermi/GBM significantly. We therefore conclude that our result is robust to differences in the assumed broadband spectral shape.

We note that the overall light curves in both instruments do not look stationary, thus the effects of the non-stationarity on our inferences are a concern. We checked this in two different ways. First, non-stationarity tends to lead to powers at low frequencies that are not distributed as the expected $\chi^2$ distribution with $2MW$ degrees of freedom, where $M$ corresponds to the number of averaged spectra, and $W$ to the number of averaged frequency bins \citep{vanderklis1989}. We find the residuals $R_j = 2I_j/S_j$ of pow ers $I_j$ and broadband model $S_j$ to be distributed 
overall as expected, with a mean of $2.0$ and a variance of $4/\sqrt{MW} = 1.633$, where $M=3$ and $W=1$ in this case. At low frequencies, where deviations are expected, a Kolmogorov-Smirnov test between the residuals and the appropriately scaled $\chi^2$ distribution yields a p-value of $0.02$. While this is on the low end, we note that this approach does not take the uncertainty in the broadband modeling into account: some powers might deviate from the expected distributions purely because we did not manage to find the maximum likelihood estimate exactly. Thus, we conclude that there is not sufficient evidence to reject the null hypothesis that the powers are $\chi^2$ distributed. 
As a second test, we also constructed two end-matched segments from each of the observations in order to minimize the effect of overall trends and red noise leak, and consider the power spectra of these segments individually. We find excess power in all end-matched sub-selections of the light curve at the frequency of the QPO. While none is significant in its own right, given the large number of trials, they do account, when averaged, for the strength of the observed QPO in both instruments. We therefore do not consider non-stationarity an important factor on our detection significance for either \swift/XRT or \fermi/GBM observation considered here.

The fractional rms amplitude is $r_{\mathrm{frac}} = 0.18\pm 0.02$ and $r_{\mathrm{frac}} = 0.033\pm 0.004$ for \swift/XRT and \fermi/GBM, respectively.
Using Fisher's method \citep{fisher1925}, we combine the two p-values (assuming mutual independence) to the test statistic $T_p = -2\sum_{k=1}^{K}{\log{p_k}}$, which is distributed following a $\chi^2$ distribution with $2K$ degrees of freedom, where $K$ is the number of independent p-values included in the analysis. 
We find a combined p-value of finding the two detections in independent, but simultaneous data sets to be $p = 4.6\times 10^{-3}$. Note that this is an upper limit on the p-value. The Bayesian method we employ computes the p-value of discovering a significant outlier in the entire power spectrum considered. As such, the p-value quoted here represents the probability of observing an outlier in two simultaneous data sets if there is no signal in either data set, irrespective of the signal's frequency. The true p-value will be smaller due to the fact that we do not only see the QPO in both \swift\ and \fermi\ simultaneously, but also at the same frequency. 

\begin{figure*}[htbp]
\begin{center}
\includegraphics[width=\textwidth]{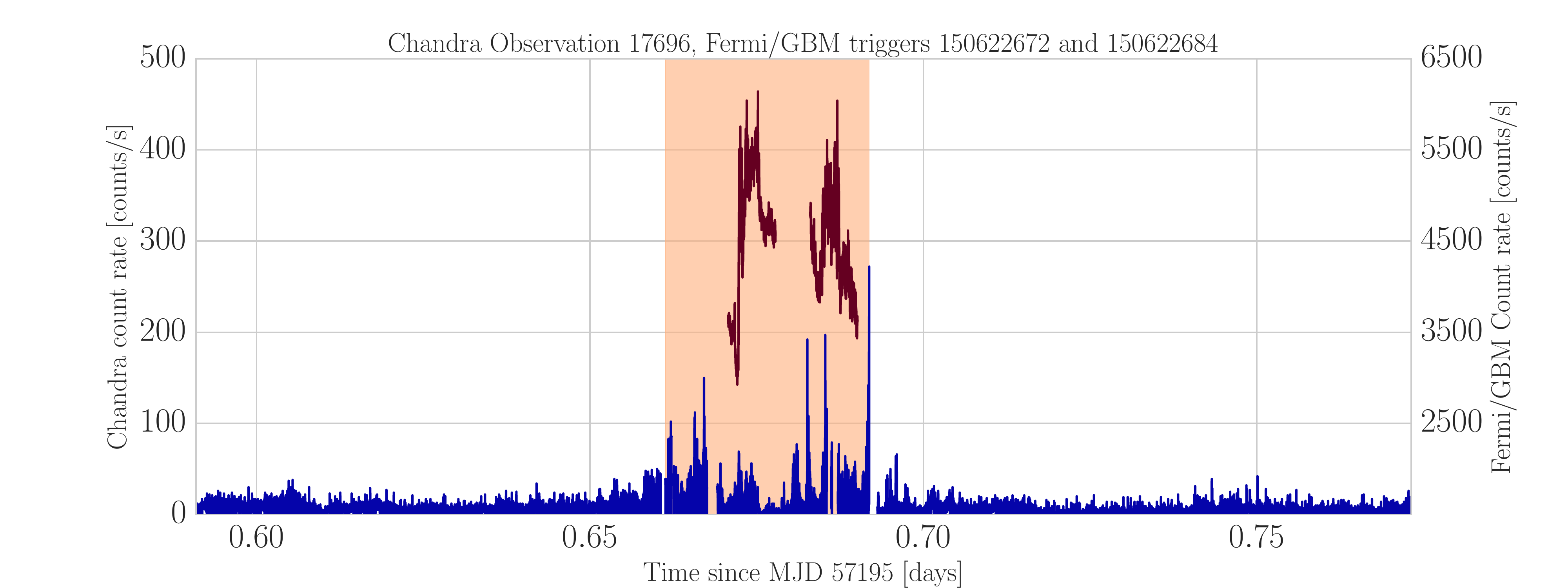}
\caption{Light curve of the first half of \chandra\ observation $17696$ (blue) and simultaneous \fermi/GBM triggers $150622672$ and $150622684$ (red). The part of the observation containing the QPOs at $73\,\mathrm{mHz}$ and $1.03\,\mathrm{Hz}$ are bounded by the orange rectangle.}
\label{fig:chandra_qpo_lcs}
\end{center}
\end{figure*}
\begin{figure*}[htbp]
\begin{center}
\includegraphics[width=\textwidth]{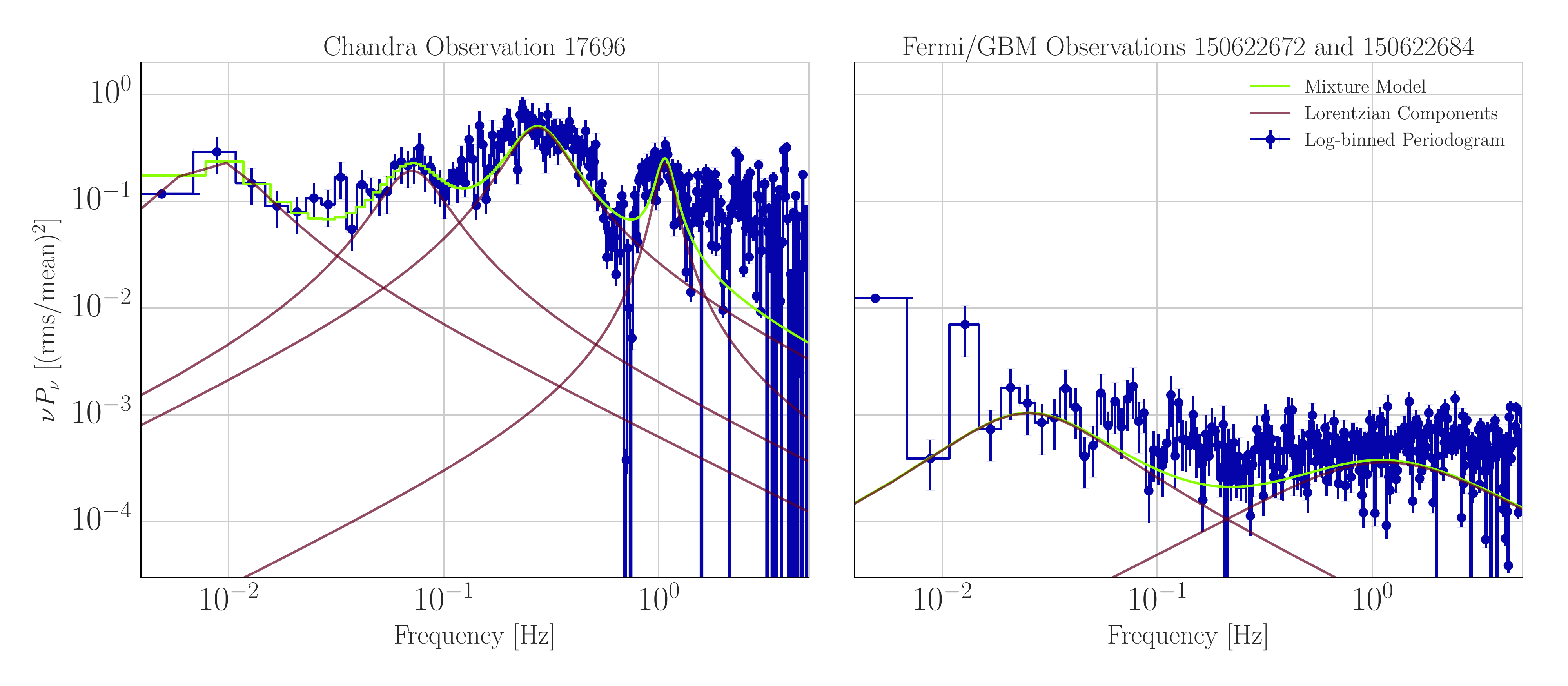}
\caption{Left panel: averaged periodogram of the part of \chandra\ observation $17696$ containing the QPOs at $73\,\mathrm{mHz}$ and $1.03\,\mathrm{Hz}$ (left panel) and the averaged periodogram of the two \fermi/GBM triggers simultaneous with the \chandra\ data (right panel). In blue, we show the  logarithmically binned periodogram.
For both data sets, we show the MAP model with four (\chandra) or two (\fermi/GBM) Lorentzian components in purple and the combined model in green. In the \chandra\ observations, two Lorentzians model QPOs, and two model the broad-band noise components. In the \fermi/GBM data set, there is no QPO present, and the two Lorentzians model broadband noise components only. The constant component modelling the Poisson level is not shown.}
\label{fig:chandra_qpo}
\end{center}
\end{figure*}

\begin{figure*}[htbp]
\begin{center}
\includegraphics[width=\textwidth]{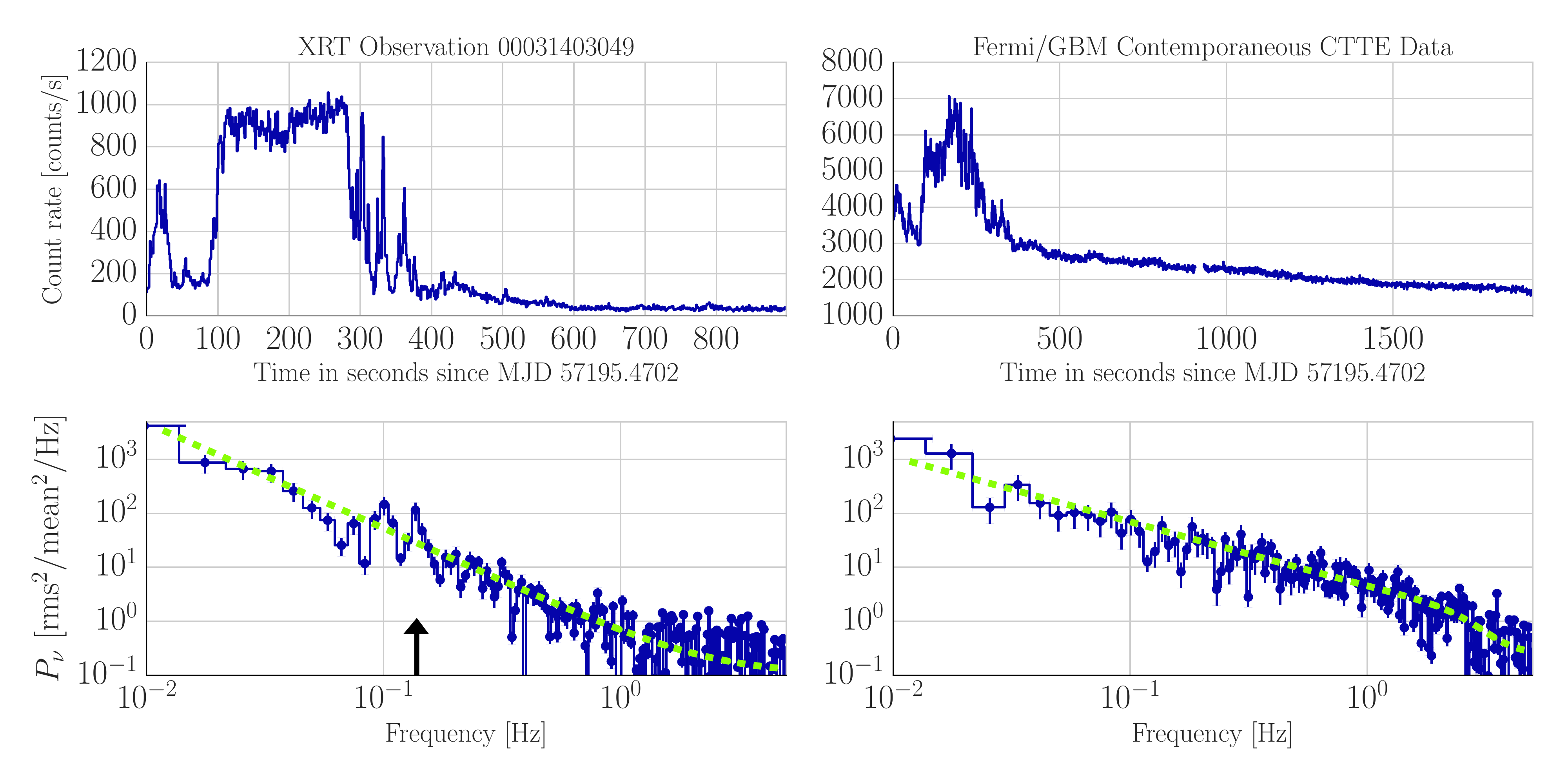}
\caption{Top panels: Light curves of \swift/XRT observation $00031403049$ (left) and the simultaneous \fermi/GBM CTTE data (right). Bottom panels: Averaged periodograms (blue data points) for the same observations, showing a QPO in the \swift/XRT observation at $136\,\mathrm{mHz}$ (black arrow). The green dashed line represents the MAP model of the broadband variability. Most of the QPO variability is concentrated in the first $400\,\mathrm{s}$ of the light curve. In particular, there is very regular QPO-like behaviour visible in the decay between $270\,\mathrm{s}$ and $450\,\mathrm{s}$ at the frequency where the QPO is detected. However, using the averaged power spectrum of the full light curve increases the strength of the signal significantly.} We note that the second peak at $97\,\mathrm{mHz}$ is not significant (posterior predictive p-value: $p = 0.18 \pm 0.01$).
\label{fig:xrt_qpo2}
\end{center}
\end{figure*}

Furthermore, we report the detection of two QPOs in a part of the \chandra/ACIS observation $17696$, starting at $\mathrm{MJD} = 57195.669$ (see Figure \ref{fig:chandra_qpo_lcs} for a light curve). The part of the observation with QPOs is simultaneous with \fermi/GBM triggers $150622672$ and $150622684$ (also Figure \ref{fig:chandra_qpo_lcs}), although no similar detection is made in the latter two observations. The QPOs in the \chandra/ACIS data are clearly visible in the periodogram, but too broad to be easily characterized by a p-value measuring significance of an outlier of a single power under the null hypothesis.
Instead of determining its significance that way, we opt for an approach optimizing the posterior of a model that is a mixture of two Lorentzian components for the broadband variability and an additional two Lorentzians describing the QPOs. 
We compare the BIC for a model with all four components to models with each of the Lorentzians modeling the QPOs removed, and find $\Delta\mathrm{BIC} = 9.97$ for the lower-frequency QPO and $\Delta\mathrm{BIC} = 29.79$ for the higher-frequency QPO, indicating very strong evidence against the model that excludes each QPO, compared to a model including a Lorentzian for that QPO.
A similar model comparison for the simultaneous \fermi/GBM triggers gives $\Delta\mathrm{BIC} = -12.13$ and $\Delta\mathrm{BIC} = -16.195$, indicating that here, the model without QPOs is strongly favoured and neither component is present in \fermi/GBM. It is unclear whether the non-detection in \fermi/GBM indicates a strong energy-dependence of the QPO or an instrumental effect. \fermi/GBM is much less sensitive, afflicted by a much higher background, and the duration of the two triggers is shorter than of the \chandra\ light curve used. It may well be the case that the lower quality of the resulting data is sufficient in explaining the discrepancies between the two power spectra.
The averaged periodograms of the relevant \chandra/ACIS and simultaneous \fermi/GBM light curves are shown in Figure \ref{fig:chandra_qpo}. The lower-frequency QPO has a centroid frequency of $\nu_{\mathrm{c}} = 73\,\mathrm{mHz}$, a width of $12\,\mathrm{mHz}$   and a fractional rms amplitude of $0.27 \pm 0.03$. The higher-frequency QPO has a  centroid frequency of $\nu_{\mathrm{c}} = 1.03\,\mathrm{Hz}$, a width of $0.11 \,\mathrm{Hz}$ and a fractional rms amplitude of $0.46 \pm 0.02$. 

We computed the fractional rms amplitude by integrating over the 
Lorentzian component modelling this QPO and estimated the error using Monte Carlo simulations of the entire mixture model using the inverse Hessian to estimate the error in the parameter estimates. We integrated over $1000$ model components at the same QPO frequency 
as the best-fit model and computed the standard deviation of the fractional rms amplitude in each simulation. No QPO is found in the 
Chandra data either before or after the light curve in question, indicating that the signal might be transient.

The corresponding upper limits for the \fermi/GBM data at the detection frequencies of the \chandra\ QPOs are $0.008$ at $73\,\mathrm{mHz}$ and $0.004$ at $1.03\,\mathrm{Hz}$. It is important to keep in mind, however, that these upper limits were derived for a single frequency, whereas the fractional rms amplitudes for the \chandra\ detections were derived by integrating over the Lorentzian model used to represent these features.

Because the observations were taken in an unusual observing mode, possible instrumental effects are a concern with the two \chandra\ detections. In particular, it is in principle possible that some of the dithering signal might have leaked into the light curve with the QPO, though the segment too short to see the dithering itself in the periodogram. 
In order to check this, we compared the periodogram with the QPO to that of a long segment of $22~\mathrm{ks}$ from \chandra/ACIS observation $17697$. Using the MEG data, where the dithering signal is exceptionally prominent and visible to the eye, we produce a periodogram of this segment and clearly identify a signal at $1.45\,\mathrm{mHz}$ (the dithering frequency). In comparison, the dithering signal is largely absent in the HEG data used for the QPO search, with the peak reduced by a factor of $6$ and comparable to the broadband noise at this frequency. 
At the frequencies of the detected QPOs, however, the periodogram of the long MEG light curve is very clearly dominated by broadband noise power, indicating that it is unlikely that any of the observed QPO power is due to instrumental effects related to the dithering, which should be clearly observed in the MEG data if this were the case.

We note that while the \integral\ observations overlap partially with the \chandra\ data in time, they end shortly before the appearance of the QPOs in the \chandra\ data. 
We find no credible detection in the simultaneous data observed simultaneously with \chandra\ and \integral, despite a relatively high signal-to-noise ratio in both and the long duration of the observations.

\begin{table*}[htb]
\renewcommand{\arraystretch}{1.3}
\footnotesize
\caption{Overview of the QPO detections}
\begin{threeparttable} 
\begin{tabularx}{18cm}{p{2.3cm}p{2.3cm}p{2.3cm}p{2.3cm}p{2.3cm}p{2.3cm}p{2.3cm}}
\toprule
\bf{MJD}        & \bf{Instrument}  & \bf{QPO frequency}      & \bf{$q$-factor} & \bf{QPO fractional rms amplitude} & \bf{$p$-value}\tnote{\emph{a}} & \bf{$\Delta$BIC} \\ \midrule
57195.47033 & \swift/XRT         & $18 \,\mathrm{mHz}$   & $\sim 4.5$        & $0.18\pm 0.02$  & $0.02$ & \\
57195.47033 & \fermi/GBM       & $18 \,\mathrm{mHz}$   & $\sim 4.5$        & $0.03\pm 0.01$  
& $0.02$ & \\
57195.66909 & \chandra/ACIS  & $73 \,\mathrm{mHz}$   & $\sim 6.0$                &  $0.27\pm 0.03$ & & $9.97$ \\
57195.66909 & \chandra/ACIS  & $1.03 \,\mathrm{Hz}$   & $9.0$                & $0.46 \pm 0.02$ & & $29.79$ \\
57195.47244 & \swift/XRT       & $136 \,\mathrm{mHz}$ & $\sim 5.8$         &  $0.08\pm 0.02$ 
& $1.9\times 10^{-3}$ & \\
\bottomrule
\end{tabularx}
   \begin{tablenotes}
      \item[\emph{a}]{Posterior predictive $p$-value for a single QPO detection, as described in the text.}
\end{tablenotes}
\end{threeparttable}
\label{tab:qpos}
\end{table*}

Finally, we report a detection in \swift/XRT data at $136\,\mathrm{mHz}$ (see Figure \ref{fig:xrt_qpo2}) with a classical p-value of $1.6\times 10^{-6}$ and a posterior predictive p-value of $1.9 \pm 0.44 \times 10^{-3}$. This detection occurs in the first orbit of observation $00031403049$, but is not present in the second. As with the detections at $18\,\mathrm{mHz}$, we produced two end-matched sub-selections of the light curve and checked whether non-stationarity might have caused the observed signals. We find that while most of the signal is concentrated in the first $400\,\mathrm{s}$ of this observation, constructing an averaged power spectrum that utilizes the full light curve increases signal strength significantly (from a classical p-value of $1\times 10^{-4}$ to $1.6\times 10^{-6}$). There is no evidence that red noise leakage has significantly affected out results. We also note that the QPO is visible by eye in the decaying part of the of the light curve between $\sim 280\,\mathrm{s}$ and $400\,\mathrm{s}$. The signal has a fractional rms amplitude of $r_\mathrm{frac} = 0.08 \pm 0.02$. There is a second feature at $98 \,\mathrm{mHz}$ in the same power spectrum, but the latter is not significant. No similar feature is observed in either the \fermi/GBM or \integral\ data sets, and it is once again unclear whether this indicates an energy dependence of a signal or should be taken as an indication that this QPO might not be of physical origin in the source. 
An overview of all QPOs detected in the data sets can be found in Table \ref{tab:qpos}.

We do not reproduce the detections at $1.8 \,\mathrm{Hz}$  and at $1.7\,\mathrm{Hz}$ in \swift/XRT observations $00031403038$ and $00644520000$, respectively, claimed in \citet{mottaatel2015, radhika2016}. We find no signal in observation $00031403038$, and an excess of power in $00644520000$, though at $2.01\,\mathrm{Hz}$ rather than 
$1.8 \,\mathrm{Hz}$. Since the trial-corrected $p$-value of this excess is merely $p = 0.025$ and there is no confirmation from another instrument, we are disinclined to claim this as a detection.

No credible QPO detections are made in either data sets from \integral\, nor in the \nustar\ data. In particular, we do not reproduce the QPO reported in \citet{prosvetov2015} in the \integral/IBIS data. This QPO is due to the fact that the instrument telemetry restart is synchronized with an 8 second frame, leading to a signal at $\sim 0.125\,\mathrm{Hz}$. The data preparation described in Section \ref{section:obs} automatically corrects for this effect, thus our data are unaffected.

\begin{figure*}[!t]
\begin{center}
\includegraphics[width=\textwidth]{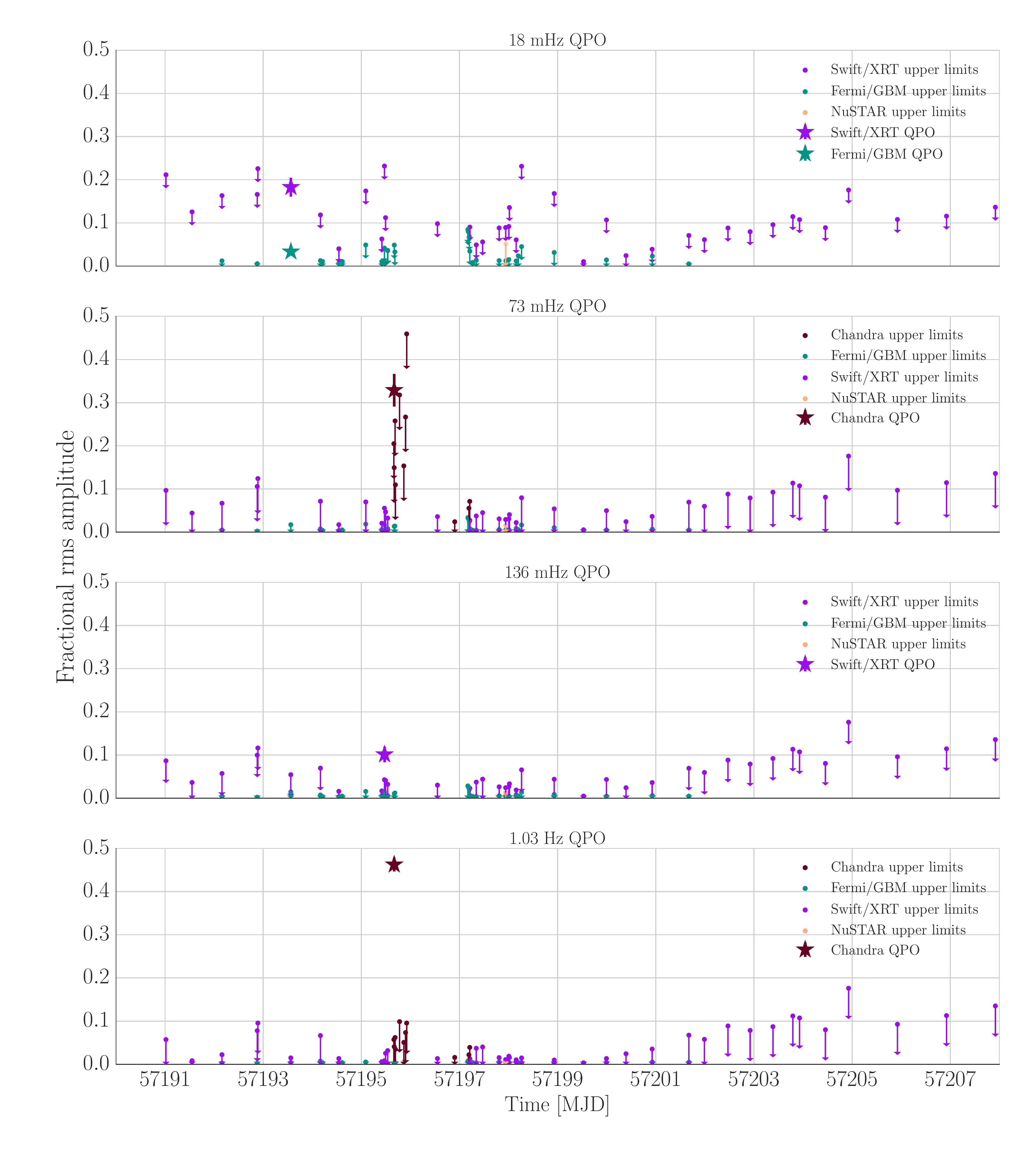}
\caption{Overview of the QPO detections as well as sensitivities for observations without detections. \integral\ observations are not included, 
since fractional rms values derived from these data are unreliable (see discussion in the text). Top panel: the QPO detected in \swift/XRT and \fermi/GBM at $18\,\mathrm{mHz}$. The fractional rms amplitudes are shown in bold stars with error bars. All other values are sensitivities at the same frequency where the QPO was observed. Middle panels : the QPO detected in \chandra\ at $73 \,\mathrm{mHz}$ and $1.03\,\mathrm{Hz}$ (pink). Sensitivities are included for the \fermi/GBM data (orange), \swift/XRT (blue) as well as the later \nustar\ observation (green). They describe the smallest fractional rms amplitude at that frequency that could have been detected in each observation. Bottom panel: the tentative QPO detection in \swift/XRT at $136\,\mathrm{mHz}$ (blue). Also included are \fermi/GBM (orange) and \nustar\ (green) sensitivities.}
\label{fig:rms_overview}
\end{center}
\end{figure*}

In Figure \ref{fig:rms_overview}, we show an overview of the fractional rms amplitudes of all detected QPOs as well as sensitivities at the relevant frequencies for observations with no detection. Since the long observations made with \chandra\ and \nustar\ are highly variable and contain strong flaring episodes, we first compute segments of $256\,\mathrm{s}$ duration and compute the variance in each segment. We then compute the median variance from all segments, and exclude all segments where the variance exceeds five times the median variance: $\sigma_{\mathrm{seg}}^2 \geq 5\sigma_{\mathrm{median}}^2$. This ensures that segments with strong flaring are excluded from the analysis. We then derive sensitivities from averaged periodograms of the remaining segments.

Particularly the two QPOs in \chandra\ are detected at a high fractional rms amplitude and with high fidelity. 
At the same time, sensitivities in the remainder of the outburst that are lower by a factor of about 5 in both \swift/XRT, operating at the same energy range, and at higher energies in \fermi/GBM indicate that the signal is either not present during the whole outburst, or very weak outside the observed \chandra\ interval. The single upper limit derived from a long, high signal-to-noise \nustar\ observation strengthens this conclusion. Similarly, the tentative detection at $136\,\mathrm{mHz}$, seen only in \swift/XRT, seems to be transient as well. 

\begin{figure*}[!t]
\begin{center}
\includegraphics[width=\textwidth]{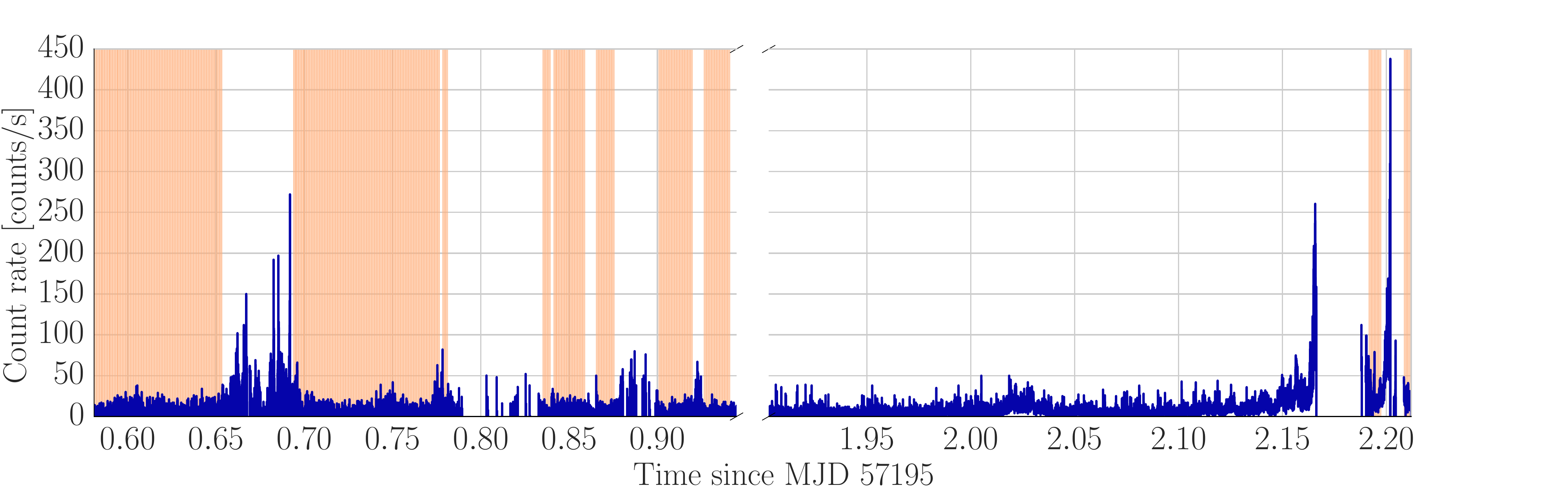}
\caption{The light curves for both \chandra\ observations $17696$ and $17697$. The intervals used for the periodogram in Figure \ref{fig:chandra_broadband} are marked in orange. We excluded intervals of bright flaring, and we also excluded the first GTI in observation $17697$, as it showed a very strong dithering signature.}
\label{fig:chandra_broadband_lc}
\end{center}
\end{figure*}

\begin{figure}[!t]
\begin{center}
\includegraphics[width=9cm]{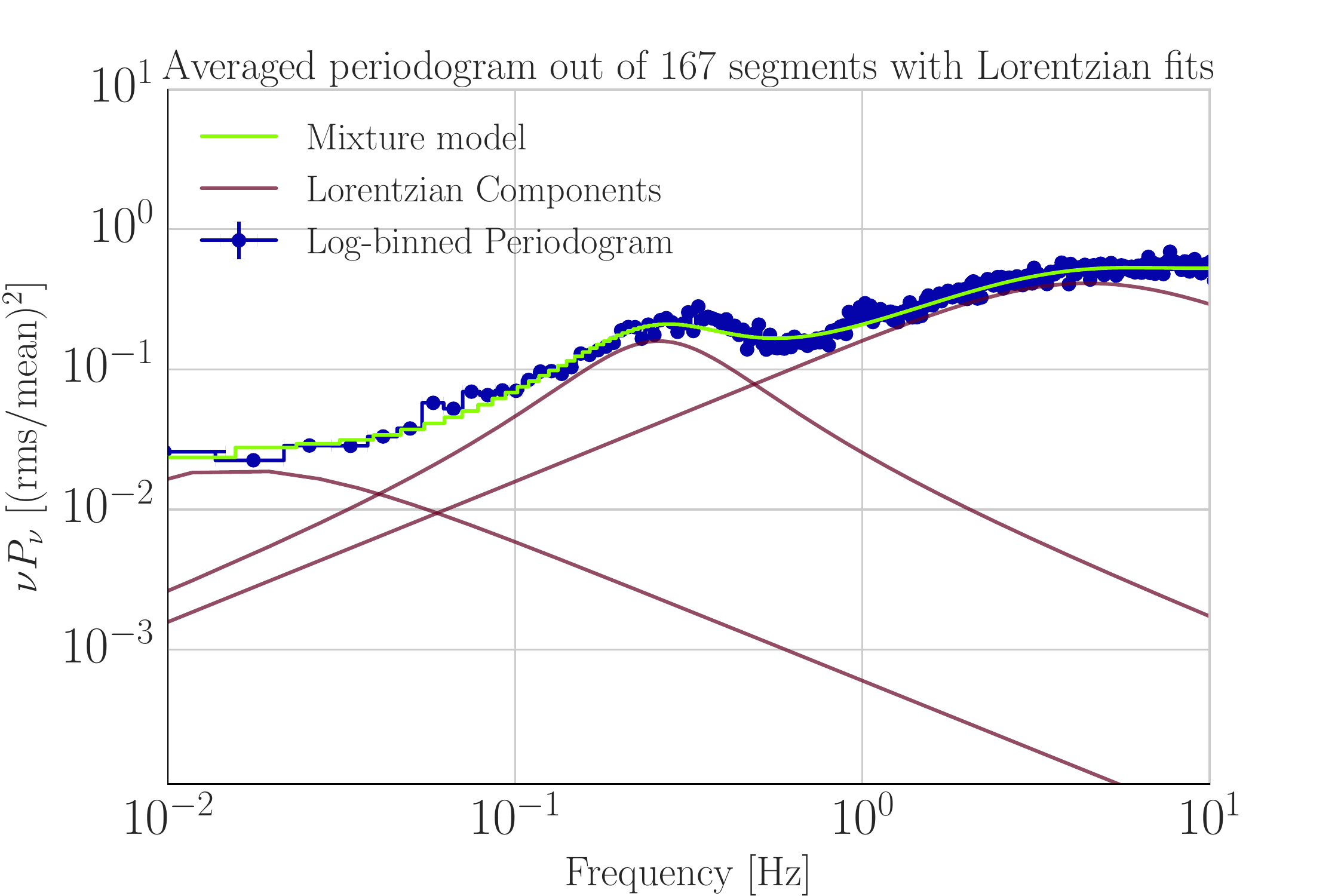}
\caption{Averaged and logarithmically binned periodogram of both \chandra\ observations (blue), along with the MAP model (green) with its individual Lorentzian components (dark red). A constant for the high-frequency Poisson noise was part of the model, but is not shown here.}
\label{fig:chandra_broadband}
\end{center}
\end{figure}

There are some observations with \swift\ very early as well as very late in the outburst that do not provide the sufficient statistics to exclude a presence of a QPO at the same frequency. 
However, the data between 2015-06-21 and 2015-06-29 are adequately constraining to conclude the signal must be short-lived or have a highly variable amplitude. 

The situation is less straightforward for the lowest-frequency QPO at $18\,\mathrm{mHz}$. Here, the short cadence of the \swift/XRT 
observations makes it hard to derive constraining sensitivities on a frequency this low. There seem to be at least parts of the outburst 
where either the QPO must vanish or its fractional rms amplitude fall below $0.1$ in the lower energy band covered by \swift.

\subsection{Broadband Variability}

In order to characterize the broadband variability in the long observations taken with \chandra, \integral\ and \nustar, we first excluded the flaring episodes as described in Section \ref{section:qposearch} by dividing the original light curve into segments of $128\,\mathrm{s}$ duration and excluding all segments for which the variance exceeds five times the median variance over all light curves. We then computed an averaged periodogram over the remaining segments for each instrument.

While the broadband variability in all observations with \swift/XRT and \fermi/GBM is adequately modelled with a broken power law, the 
\chandra\ observations require additional components (see Figure \ref{fig:chandra_broadband_lc} for the light curve and Figure \ref{fig:chandra_broadband} for the averaged periodogram). 
Aside from the two QPOs in specific parts of observation $17696$, the averaged spectrum from all available \chandra\ light curves can be well-modelled with a mixture of three Lorentzian components (denoted below as low-, mid- and high-frequency components) and a constant for the high-frequency noise. 
The first (zero-centered) Lorentzian component models the very lowest-frequency (band-limited) noise, whereas the second and third Lorentzian are required to account for structure at higher frequencies. 

Of particular interest is the strong variability component between $0.1$ and $1.0\,\mathrm{Hz}$, modeled by the mid-frequency Lorentzian with a centroid frequency of $\nu_0 = 0.24 \pm 0.01 \,\mathrm{Hz}$ and a HWHM of $\Delta\nu = 0.12\,\mathrm{Hz}$, just at the lower edge between being formally called a QPO versus broadband noise, but nevertheless clearly a visible peak in the power spectrum (see Figure \ref{fig:chandra_broadband}). 
The high-frequency component, for comparison, has a centroid frequency of $0.51 \pm 0.08\,\mathrm{Hz}$ and a HWHM of $\Delta\nu = 4.35\,\mathrm{Hz}$. 

A similar MAP fit of the averaged periodogram of the \integral/JEM-X data reveals that the same number of Lorentzian components is strongly favoured over a model with fewer components ($\Delta BIC = 40.17$ in favour of the more complex model), however, the Lorentzians are much broader and flatter in the latter data set, leading to a much smoother power spectrum. Additionally, the peak frequencies of the mid- and high-frequency Lorentzians are lower than for the \chandra\ observations: $\nu_0 = 0.011 \pm 0.007 \,\mathrm{Hz}$ (HWHM: $\Delta\nu = 0.06\,\mathrm{Hz}$) and $\nu_0 = 0.16 \pm 0.11\,\mathrm{Hz}$ (HWHM: $\Delta\nu = 0.85$) respectively, and less well constrained in general, though the difference in the centroid frequencies between \chandra\ and \integral/JEM-X are significant.
The \integral/IBIS data requires only two components at the lowest frequencies, indicating that there is less variability at higher energies. Note that because of the non-trivial statistical properties of the \integral\ data, conclusions derived from averaged periodograms should be taken with a grain of salt. 
The cospectrum averaging all non-flaring \nustar\ light curves from observation $90102007002$ is very smooth, modelled adequately by two Lorentzian components which both extend over more than an order of magnitude in width, indicating that the additional power in the \chandra\ observations in this frequency band is not present here. 

The broadband noise observations are broadly consistent with results from V404 Cygni's previous outburst in 1989 \citep{oosterbroek1997}, which consisted largely of smooth broadband noise spectra modelled by three Lorentzian components (though one component is at higher frequencies than we consider here) and saw an additional increase in power 
when the source was very bright similar to the excess observed here, but at a lower frequency of $\sim 50\,\mathrm{mHz}$. 

\section{Discussion}
\label{section:discussion}

Even though the 2015 outburst of V404 Cygni  was spectacular in both its rarity and its brightness, it actually shows comparatively little complex variability behaviour even in the states where one would traditionally expect strong broadband noise and QPOs. 
Here, we for the first time find strong evidence for four significant signals in V404 Cygni.

The QPOs in the 2015 outburst occur at $18\,\mathrm{mHz}$ in both \swift/XRT (fractional rms amplitude $r_{\mathrm{frac}} = 0.18\pm 0.02$) and \fermi/GBM ($r_{\mathrm{frac}} = 0.03\pm 0.01$), at $73\,\mathrm{mHz}$ in \chandra/ACIS ($r_{\mathrm{frac}} = 0.27 \pm 0.03$), $136\,\mathrm{mHz}$ in \swift/XRT  ($r_{\mathrm{frac}} = 0.08\pm 0.02$), and $1.03\,\mathrm{Hz}$ in \chandra/ACIS  ($r_{\mathrm{frac}} = 0.46\pm 0.02$). All signals are at relatively high fractional rms amplitude and seem to occur transiently in only a short interval during the outburst. 

Among the phenomenology of QPOs in black hole X-ray binaries, generally two classes can be distinguished: high-frequency QPOs (HFQPOs) in the range of \SEM{$100-500\,\mathrm{Hz}$} \citep[e.g.][]{remillard1999a,remillard1999b,miller2001,strohmayer2001} and low-frequency QPOs (LFQPOs) between $0.05$ and \SEM{$30\,\mathrm{Hz}$} \citep{motch1983,miyamoto1991,takizawa1997,motta2015}.

The latter category can furthermore be subdivided into types A, B and C \citep{wijnands1999,sobczak2000,lin2000,homan2001,remillard2002}. 
Type A QPOs are very broad at a low amplitude and seen during the intermediate state around a centroid frequency of $6\,\mathrm{Hz}$. This is clearly much  higher than any of the QPOs reported here. 
Similarly, Type-B QPOs also appear at frequencies of $\sim 6\,\mathrm{Hz}$, though narrower, thus none of the QPOs observed here fall into this category. Type-C QPOs, on the other hand, mostly occur in the intermediate and hard states at frequencies between $0.1\,\mathrm{Hz}$ and $30\,\mathrm{Hz}$, in reasonably good agreement with the QPOs observed here at $73\,\mathrm{mHz}$, $136\,\mathrm{mHz}$ and $1.03\,\mathrm{Hz}$. 
Additionally, given V404 Cygni's orbital inclination of $67^{+3}_{-1}$\deg \citep{shahbaz1994,khargharia2010}, it follows the general trend of Type-C QPOs in high-inclination systems to have higher fractional rms amplitudes \citep{heil2015,motta2015}, though unlike most Type-C QPOs, these QPOs are likely highly transient. 

A large fraction of the extreme variability of V404 Cygni was partly due to large changes of column density local to the source, as already seen in 1989 \citep{Oosterbroek1996,zycki1999}. However, \cite{rodriguez2015} have shown that at least part of such variability was instead intrinsic to the source (thus related to mere accretion events) and somewhat similar to that typical GRS 1915+105. This bright and highly variable system is known to display fast state transitions where the disk truncation radius varies by several tens of gravitational radii in matters of seconds \citep{belloni1997b}. If this is the case also for V404 Cyg, then the source progressed from a system resembling an advection-dominated accretion flow (ADAF) to a very luminous state where it accreted close to the Eddington limit repeatedly in a matter of hours, subsequently switching several times between a hard state and a highly luminous state. 
In this context, with luminosity changes on timescales of minutes to hours, it is unsurprising that we observe QPOs for only short periods of time, before the source moves out of a spectral regime where they are likely to be seen.

Furthermore, a rapid evolution of either the truncation radius or the radius of a ring where the anisotropies occur \citep{ingram2014} would provide a natural explanation for why the observed QPOs are short-lived and why no standard Type-C QPOs at higher frequencies between $1-15\,\mathrm{Hz}$ are observed. As the disk rapidly fills, the resulting accretion flow might be much more turbulent than it would otherwise be. In particular, a precessing flow producing a Type-C QPO (as proposed by e.g.\ \citealt{stella1998,schnittman2006,ingram2011}) requires a sound crossing time scale faster than the precession timescale \citep{ingram2009}, such that the warping of the flow is preserved during precession \citep{lubow2002,fragile2007}. In the rapidly changing flow of V404 Cygni, physical properties of the plasma such as pressure, temperature, density and viscosity might evolve on short time scales \citep{jenke2016}, leading to a highly variable sound speed and thus a range of sound crossing time scales that support precession for only short intervals. 
For example, both the QPOs observed in $00031403040$ by Swift at $18\,\mathrm{mHz}$ and that seen by \chandra\ occur during a part of the outburst when the source was in a low-luminosity state, rising toward a high-flux state, indicating that the source was quickly moving away from an accretion regime where type-C QPOs could form and survive. 
Of course, our general picture is that black hole accretion disks truncates at low luminosities \citep{esin1997,tomsick2009}, with truncation radii becoming smaller and smaller as the luminosity increases. However, it must be noted that the exact luminosity when truncation occurs is unclear, and there are examples where the accretion disk extends close to the ISCO also in the hard state \citep{miller2015, parker2015}).

Unlike the other three reported signals, the QPO detected in \swift/XRT observation $00031403049$ at $136\,\mathrm{mHz}$ is only observed in orbit $3$, where the source was extremely bright, variable and therefore in a spectral state likely closer to a luminous soft-intermediate state (or a Ultra-Luminous state, \citet{Belloni2016}) rather than a hard one. While Type-C QPOs are indeeed observed in very luminous states in BHXRBs, they are usually seen at a much higher frequency of $\sim\! 30 \,\mathrm{Hz}$ \citep{motta2012}. There is currently no consistent model to explain a QPO at a frequency this low in a state where the source is extremely bright.

While consistent with spectral states where QPOs exist, the QPO at $18\,\mathrm{mHz}$ is at too low a frequency to be readily interpreted within the standard types of LFQPOs in black hole X-ray binary systems. 
Instead, it seems to represent another instance of a growing class of mHz QPOs now observed in several black hole sources with different properties from typical Type-C QPOs.
A mHz QPO in a black hole LMXB was first observed in H1743-322 at the beginning of two outbursts in both \rxte\ and \chandra\ in 2010 and 2011 \citep{altamirano2012}. 
The QPO vanished within a few days for both occurrences, but was consistent in frequency across outbursts, indicating a stable underlying timescale different from the generally more variable Type-C QPOs simultaneously present in the same observations. 
Notably, the QPO fractional rms amplitude was stable with photon energy, again unlike Type-C QPOs, which show a marked dependence of fractional rms amplitude on energy. 
Since this initial detection, similar QPOs have been found in at least four more sources: LMC X-1 \citep{alam2014}, IC10 X-1 \citep{pasham2013}, Cygnus X-3 \citep{koljonen2011} and Swift J1357.2-0933 \citep{armaspadilla2014}. 
All sources share similar QPO properties: a relatively large fractional rms amplitude of up to 10\%, a short QPO lifetime of at most a few days and occurrence during the low-hard state (LHS). 
This distinguishes them from the ``heart beat'' QPOs observed in GRS 1915+105 \citep{belloni2000} and IGR J17091-3624 \citep{altamirano2011}, which show large-amplitude oscillations during the high soft state generally attributed to a limit cycle behaviour of a radiation pressure instability that causes quasi-periodic evaporation of the inner parts of the accretion disk followed by a refilling of the same \citep{lightman1974,belloni1997a,neilsen2011}.

In the context of the detection of a QPO at a similarly low frequency in V404 Cygni, also transient and occurring near the beginning of the outburst, there is another important commonality four of the other five systems share: they show either optical or X-ray dips, which led to the hypothesis that these QPOs could be analogous to the $1\,\mathrm{Hz}$ QPO seen in some dipping neutron star XRBs, thought to be either due to an accretion disk structure obscuring the inner hot region or to relativistic Lense-Thirring precession of the inner accretion disk \citep{homan2012}.

The dipping neutron star systems where the $1\,\mathrm{Hz}$ QPO is observed are believed to be systems with a high orbital inclination, and indeed, this explanation is only feasible if the system is seen nearly edge-on, where periodic obscuration might be visible as a QPO-like feature in the light curve. 
H1743-322, LMC X-1, IC10 X-1 and Swift J1357.2-0933 are all believed to be at a high orbital inclination, whereas this is not true for Cygnus X-3. 
In Cygnus X-3, a high-mass X-ray binary with a Wolf-Rayet companion \citep{vankerkwijk1992}, the QPO detections followed major radio flares \citep{koljonen2011}, indicating that the QPO and the jet are linked in this system, with either the jet shadowing oscillatory behaviour in the corona or with the QPOs caused by a structure in the jet itself.
Unlike Cygnus X-3, V404 Cygni's relatively high orbital inclination could add it to the growing sample of high-inclination sources that show these mHz QPOs, though no dips have been observed from the source. 
Much like the other four edge-on systems, the mHz QPO in V404 Cygni is observed near the start of the outburst, although not as close to the start as in H1743-322, and it was likely short-lived (duration $<1\,\mathrm{day}$). 
In contrast to other mHz QPOs, however, there seems to be a strong dependence of fractional rms amplitude on energy: the QPO is significantly stronger at lower photon energies in \swift/XRT than it is at high energies in \fermi/GBM. 
It is unclear whether this could be due to the intrinsic differences in sensitivity, background and instrument collecting area. 

If the association of the low-frequency QPO observed here with the $1\,\mathrm{Hz}$-QPO observed in dipping neutron star systems were true, then the difference in frequency might either be explained if the QPO scales with mass, or if it depends on the orbital period of the system, though the latter was ruled out by the short orbital period of Swift J1357.2-0933 \citep{armaspadilla2014}. Similarly for V404 Cygni we can rule out mass scaling: in order to reduce the precession frequency from $1 \,\mathrm{Hz}$ to $18 \,\mathrm{mHz}$, the mass of the black hole would have to be $\sim 77 \,\mathrm{M}_\odot$, inconsistent with previous mass estimates.

Relativistic precession could still occur if either the truncation radius is relatively large or precession occurs further out in the disk. Again, the peculiarities and rapid state evolution may play an important role in explaining the low frequency of this QPO. 
Perhaps it is not the accretion flow that precesses, but the jet \citep[e.g.][]{kalamkar2015}. This explanation might be supported by the strong radio activity coincident with the V404 Cygni X-ray flaring \citep{mooley2015a,mooley2015b}. For blazars, relativistic beaming in a precessing jet has been proposed as an explanation for periodic flaring \citep{abraham2000, caproni2013}, though this scenario is precluded in the case of V404 Cygni by the source's high orbital inclination. 

One alternative explanation for the origin of this QPO could be warping of the outer accretion disk, either via torques exerted by the companion star on the disk \citep{tremain2014} or induced by radiation pressure \cite{pringle1996}. As the hotter inner disk illuminates the outer disk, it exerts radiation pressure upon the latter. Anisotropies in the radiation may lead to anisotropies in the radiation pressure, and consequently to perturbations that grow in a non-linear manner, and may cause warping in the disk that is responsible for the modulations in the observed X-ray flux \citep{pringle1996}. The latter scenario has been invoked to explain a low-frequency quasi-periodic oscillation in the neutron star LMXB 4U 1626$-$67 \citep{raman2016} as well as a possible driving force behind jet precession in HLX-1 \citep{king2014} and more generally as an explanation for super-orbital periods in binary systems with compact objects \citep[e.g.][]{kotze2012}.

The broad, peaked component at $190\,\mathrm{mHz}$ in the \chandra\ data is reminiscent of a similar component at $40\,\mathrm{mHz}$ observed during the 1989 outburst by \citet{oosterbroek1997}, albeit at a higher frequency. 
\citet{oosterbroek1997} likened the power spectrum, as well as that specific peak, to Cygnus X-1 observations obtained with \textit{SIGMA} onboard the \textit{GRANAT} satellite \citep{vikhlinin1994}. 
Cygnus X-1 shows an occasional peak at very similar frequencies that is too broad to be strictly interpreted as a QPO, transient and at a fairly high fractional rms amplitude when present \citep{angelini1992,kouveliotou1993,vikhlinin1994}. 
It is unclear if the broad component in the \chandra\ data obtained during the 2015 outburst can be identified with that at lower frequencies in the previous outburst. 
The behaviour of the similar feature in Cygnus X-1 would argue against that interpretation: the $40\,\mathrm{mHz}$ peak in Cygnus X-1 is remarkably stable over years of observations \citep{angelini1992}.
On the other hand, \citet{pottschmidt2003} find that the overall power spectrum of Cygnus X-1, especially during the hard state, is well-described by four peaked components modelled as a mixture of Lorentzians in the $10^{-3} - 10^{2}\,\mathrm{Hz}$ range. In particular, their middle two components $L_2$ and $L_3$ are at very similar frequencies as the upper two Lorentzians shown in Figure \ref{fig:chandra_broadband}, indicating perhaps a common origin. Similarly, \citet{axelsson2005} model the power spectrum of Cygnus X-1 with a mixture of two Lorentzians as well as a power law at the lowest frequencies, and find that the frequencies and fractional rms amplitudes of these components change during an outburst. In the hard state, Cygnus X-1 is well-modelled by two Lorentzians. As it transitions to the soft state, the Lorentzian components shift to higher frequencies and weaken until they leave behind a smooth power law.
While V404 Cygni never fully enters the soft state, we see power spectra in the \chandra\ observations that are similar to those of the intermediate states in Cygnus X-1: two moderately strong Lorentzian components as well as a Lorentzian centered on zero to describe the lowest powers (qualitatively similar to a broken power law over the frequencies of interest). On the other hand, the \nustar\ observations are adequately described by only two, very broad and fairly weak Lorentzian components, indicating a decline in overall variability.
The best explanation of the broad-band noise components observed here is provided by propagating fluctuations in the mass accretion rate \citep[e.g.][]{lyubarskii1997, uttley2005, ingram2013}. In this model, small perturbations in the mass accretion rate are propagated through the accretion disk into the inner region, where they are finally translated into fluctuations observed in the radiation. 

V404 Cygni lacks the monitoring data over long timescales in the hard state (partly due to its long quiescent intervals) to track and characterize the variability behaviour over long timescales as has been done for Cygnus X-1, GX 339-4 and other sources. The observations with \swift/XRT are too short and sparse to reliably estimate the parameters of a multi-component model and confirm whether the broadband noise changes significantly over the course of the outburst. 
The only further constraint comes from the \integral/JEM-X data immediately prior to the \chandra\ observations. The JEM-X data requires the same number of Lorentzian components to yield an acceptable fit, however, the Lorentzian centroids are at significantly lower frequencies than seen in the \chandra\ data. 
If these components are indeed the same as in the \chandra\ data, then this would imply that the broadband noise components  move to higher frequencies as the outburst progresses.  This is similar to the behaviour seen in other BHXRBs, where broadband noise components tend to shift to higher frequency as the source moves from the hard through the intermediate state into the soft state, where the variability, including the Lorentzian components, is usually strongly suppressed \citep{dimatteo1999, gilfanov1999,nowak2000,revnivtsev2000,kalemci2001,nowak2002,kalemci2003,pottschmidt2003,kalemci2005,belloni2005,kleinwolt2008,grinberg2014}.

One important caveat to the broadband variability described in this work lies in the dramatic changes the source underwent over the course of a single observations. Thus, it might be possible that spectra of several states have been averaged into the same power spectrum. If the power spectrum changes significantly between states, as is expected for a black hole XRB, it is possible that the power spectra reported here show fewer features than would otherwise be the case, since the distinctive shape of the Lorentzian components has been smeared out during the fast state transitions.

\section{Conclusions}
We present a comprehensive search for QPOs in X-ray observations of V404 Cygni during its most recent outburst in 2015 June-July. We detect for the first time QPOs in this source, at frequencies of $18\,\mathrm{mHz}$ (\swift/XRT and \fermi/GBM); $73\,\mathrm{mHz}$, $1.03\,\mathrm{Hz}$ (\chandra); $136\,\mathrm{mHz}$ (\swift/xrt). 
mHz QPOs in black hole XRBs are rare; this is only the fifth such signal. They have been observed from both LMXBs and HMXBs, with common properties slowly emerging for most of them: occurrence near the beginning of an outburst, frequencies in the $5-30\,\mathrm{mHz}$ range, a short lifetime of a few days or less. 
All but one, including the $18\,\mathrm{mHz}$ QPO detected in V404 Cygni in this work, come from sources with an inferred high inclination. In principle, this might suggest an interpretation analogous to the $1\,\mathrm{Hz}$ QPOs seen in dipping neutron stars as modulation of geometric structures in the outer accretion disk or obscuration of features in the inner accretion flow by outer parts of the disk. However, an origin in radiation pressure-drive warping of the outer disk caused by anisotropies in the radiation incident upon the disk is equally compelling. 
The remaining QPOs observed in V404 Cygni fit the general behaviour expected for type-C QPOs from a high-inclination source, though it remains surprising that these are the first QPOs detected in V404 Cygni and that these QPOs seem to be very short-lived compared to other sources. The transient nature might be related to the rapid spectral changes as the source moves between a hard state and an intermediate-soft state on timescales of hours, allowing little time for the stable precession required to induce QPOs in the X-ray radiation.
Given the data quality, it is difficult to characterize the behaviour of the broadband noise over the course of the outburst. Using the highest-quality data available from \chandra\ and \integral/JEM-X, we find the power spectrum between $0.005\,\mathrm{Hz}$ and $5\,\mathrm{Hz}$ is well-modelled by a mixture of three Lorentzians, and these Lorentzians seem to shift to higher frequencies with time, similar to previous observations of the long-term behaviour of Cygnus X-1. 

\paragraph{acknowledgements}
We thank the anonymous referee for their very helpful comments and suggestions. We thank Victoria Grinberg for very informative and useful discussions on \integral\ calibration and timing with IBIS/ISGRI.
DH acknowledges support by the Moore-Sloan Data Science Environment at NYU.  Partly based on observations with \integral, an ESA project with instruments and science data centre funded by ESA member states (especially the PI countries: Denmark, France, Germany, Italy, Switzerland, Spain) and with the participation of Russia and the USA. AI acknowledges support from the Netherlands Organization for Scientific Research (NWO) Veni Fellowship, grant number 639.041.437. MB was supported in part by the Sardinian Region, in the framework of the Regional Fundamental Research funds (L.R. 7/2007). JC thanks financial support from ESA/PRODEX Nr. 90057.

\bibliography{td}

\begin{thebibliography}{}
\expandafter\ifx\csname natexlab\endcsname\relax\def\natexlab#1{#1}\fi

\bibitem[{{Abraham}(2000)}]{abraham2000}
{Abraham}, Z. 2000, \aap, 355, 915

\bibitem[{{Alam} {et~al.}(2014){Alam}, {Dewangan}, {Belloni}, {Mukherjee}, \&
  {Jhingan}}]{alam2014}
{Alam}, M.~S., {Dewangan}, G.~C., {Belloni}, T., {Mukherjee}, D., \& {Jhingan},
  S. 2014, \mnras, 445, 4259

\bibitem[{{Altamirano} \& {Strohmayer}(2012)}]{altamirano2012}
{Altamirano}, D., \& {Strohmayer}, T. 2012, \apjl, 754, L23

\bibitem[{{Altamirano} {et~al.}(2011){Altamirano}, {Belloni}, {Linares}, {van
  der Klis}, {Wijnands}, {Curran}, {Kalamkar}, {Stiele}, {Motta},
  {Mu{\~n}oz-Darias}, {Casella}, \& {Krimm}}]{altamirano2011}
{Altamirano}, D., {Belloni}, T., {Linares}, M., {et~al.} 2011, \apjl, 742, L17

\bibitem[{{Angelini} {et~al.}(1992){Angelini}, {White}, \&
  {Stella}}]{angelini1992}
{Angelini}, L., {White}, N.~E., \& {Stella}, L. 1992, \iaucirc, 5580

\bibitem[{{Armas Padilla} {et~al.}(2014){Armas Padilla}, {Wijnands},
  {Altamirano}, {M{\'e}ndez}, {Miller}, \& {Degenaar}}]{armaspadilla2014}
{Armas Padilla}, M., {Wijnands}, R., {Altamirano}, D., {et~al.} 2014, \mnras,
  439, 3908

\bibitem[{{Axelsson} {et~al.}(2005){Axelsson}, {Borgonovo}, \&
  {Larsson}}]{axelsson2005}
{Axelsson}, M., {Borgonovo}, L., \& {Larsson}, S. 2005, \aap, 438, 999

\bibitem[{Bachetti(2015)}]{MaltPyNT}
Bachetti, M. 2015, Astrophysics Source Code Library, record ascl:1502.021

\bibitem[{Bachetti {et~al.}(2015)Bachetti, Harrison, Cook, Tomsick, Schmid,
  Grefenstette, Barret, Boggs, Christensen, Craig, Fabian, F{\"u}rst, Gandhi,
  Hailey, Kara, Maccarone, Miller, Pottschmidt, Stern, Uttley, Walton, Wilms,
  \& Zhang}]{Bachetti+15}
Bachetti, M., Harrison, F.~A., Cook, R., {et~al.} 2015, ApJ, 800, 109

\bibitem[{{Barret} \& {Vaughan}(2012)}]{barret2012}
{Barret}, D., \& {Vaughan}, S. 2012, \apj, 746, 131

\bibitem[{{Barthelmy} {et~al.}(2015){Barthelmy}, {D'Ai}, {D'Avanzo}, {Krimm},
  {Lien}, {Marshall}, {Maselli}, \& {Siegel}}]{barthelmyatel2015}
{Barthelmy}, S.~D., {D'Ai}, A., {D'Avanzo}, P., {et~al.} 2015, GRB Coordinates
  Network, 17929

\bibitem[{Belloni \& Hasinger(1990)}]{BelloniHasinger90}
Belloni, T., \& Hasinger, G. 1990, A{\&}A, 230, 103

\bibitem[{{Belloni} \& {Hasinger}(1990)}]{belloni1990}
{Belloni}, T., \& {Hasinger}, G. 1990, \aap, 227, L33

\bibitem[{{Belloni} {et~al.}(2005){Belloni}, {Homan}, {Casella}, {van der
  Klis}, {Nespoli}, {Lewin}, {Miller}, \& {M{\'e}ndez}}]{belloni2005}
{Belloni}, T., {Homan}, J., {Casella}, P., {et~al.} 2005, \aap, 440, 207

\bibitem[{{Belloni} {et~al.}(2000){Belloni}, {Klein-Wolt}, {M{\'e}ndez}, {van
  der Klis}, \& {van Paradijs}}]{belloni2000}
{Belloni}, T., {Klein-Wolt}, M., {M{\'e}ndez}, M., {van der Klis}, M., \& {van
  Paradijs}, J. 2000, \aap, 355, 271

\bibitem[{{Belloni} {et~al.}(1997{\natexlab{a}}){Belloni}, {M{\'e}ndez},
  {King}, {van der Klis}, \& {van Paradijs}}]{belloni1997b}
{Belloni}, T., {M{\'e}ndez}, M., {King}, A.~R., {van der Klis}, M., \& {van
  Paradijs}, J. 1997{\natexlab{a}}, \apjl, 488, L109

\bibitem[{{Belloni} {et~al.}(1997{\natexlab{b}}){Belloni}, {M{\'e}ndez},
  {King}, {van der Klis}, \& {van Paradijs}}]{belloni1997a}
---. 1997{\natexlab{b}}, \apjl, 479, L145

\bibitem[{{Belloni} \& {Motta}(2016)}]{Belloni2016}
{Belloni}, T.~M., \& {Motta}, S.~E. 2016, ArXiv e-prints, arXiv:1603.07872

\bibitem[{{Belloni} {et~al.}(2011){Belloni}, {Motta}, \&
  {Mu{\~n}oz-Darias}}]{belloni2011}
{Belloni}, T.~M., {Motta}, S.~E., \& {Mu{\~n}oz-Darias}, T. 2011, Bulletin of
  the Astronomical Society of India, 39, 409

\bibitem[{{Berger} \& {van der Klis}(1998)}]{berger1998}
{Berger}, M., \& {van der Klis}, M. 1998, \aap, 340, 143

\bibitem[{{Burns} {et~al.}(2015){Burns}, {Jenke}, {Mailyan}, {Younes}, \& {von
  Kienlin}}]{burns2015}
{Burns}, E., {Jenke}, P., {Mailyan}, B., {Younes}, G., \& {von Kienlin}, A.
  2015, GRB Coordinates Network, 17948

\bibitem[{{Burrows} {et~al.}(2005){Burrows}, {Hill}, {Nousek}, {Kennea},
  {Wells}, {Osborne}, {Abbey}, {Beardmore}, {Mukerjee}, {Short}, {Chincarini},
  {Campana}, {Citterio}, {Moretti}, {Pagani}, {Tagliaferri}, {Giommi},
  {Capalbi}, {Tamburelli}, {Angelini}, {Cusumano}, {Br{\"a}uninger}, {Burkert},
  \& {Hartner}}]{burrows2005}
{Burrows}, D.~N., {Hill}, J.~E., {Nousek}, J.~A., {et~al.} 2005, \ssr, 120, 165

\bibitem[{{Canizares} {et~al.}(2005){Canizares}, {Davis}, {Dewey}, {Flanagan},
  {Galton}, {Huenemoerder}, {Ishibashi}, {Markert}, {Marshall}, {McGuirk},
  {Schattenburg}, {Schulz}, {Smith}, \& {Wise}}]{canizares2005}
{Canizares}, C.~R., {Davis}, J.~E., {Dewey}, D., {et~al.} 2005, \pasp, 117,
  1144

\bibitem[{{Caproni} {et~al.}(2013){Caproni}, {Abraham}, \&
  {Monteiro}}]{caproni2013}
{Caproni}, A., {Abraham}, Z., \& {Monteiro}, H. 2013, \mnras, 428, 280

\bibitem[{{Casares} {et~al.}(1992){Casares}, {Charles}, \&
  {Naylor}}]{casares1992}
{Casares}, J., {Charles}, P.~A., \& {Naylor}, T. 1992, \nat, 355, 614

\bibitem[{{Courvoisier} {et~al.}(2003){Courvoisier}, {Walter}, {Beckmann},
  {Dean}, {Dubath}, {Hudec}, {Kretschmar}, {Mereghetti}, {Montmerle},
  {Mowlavi}, {Paltani}, {Preite Martinez}, {Produit}, {Staubert}, {Strong},
  {Swings}, {Westergaard}, {White}, {Winkler}, \&
  {Zdziarski}}]{courvoisier2003}
{Courvoisier}, T.~J.-L., {Walter}, R., {Beckmann}, V., {et~al.} 2003, \aap,
  411, L53

\bibitem[{{Di Matteo} \& {Psaltis}(1999)}]{dimatteo1999}
{Di Matteo}, T., \& {Psaltis}, D. 1999, \apjl, 526, L101

\bibitem[{{Done} {et~al.}(2007){Done}, {Gierli{\'n}ski}, \&
  {Kubota}}]{done2007}
{Done}, C., {Gierli{\'n}ski}, M., \& {Kubota}, A. 2007, \aapr, 15, 1

\bibitem[{{Esin} {et~al.}(1997){Esin}, {McClintock}, \& {Narayan}}]{esin1997}
{Esin}, A.~A., {McClintock}, J.~E., \& {Narayan}, R. 1997, \apj, 489, 865

\bibitem[{{Ferrigno} {et~al.}(2015){Ferrigno}, {Fotopoulou}, {Domingo},
  {Alfonso-Garz{\'o}n}, {Rodriguez}, {Motta}, {Kuulkers}, {Sanchez-Fernandez},
  \& {Cadolle Bel}}]{ferrigno2015}
{Ferrigno}, C., {Fotopoulou}, S., {Domingo}, A., {et~al.} 2015, The
  Astronomer's Telegram, 7662

\bibitem[{Fisher(1925)}]{fisher1925}
Fisher, R.~A. 1925, Statistical Methods for Research Workers ) (Oliver and
  Boyd, Edinburgh)

\bibitem[{{Foreman-Mackey} {et~al.}(2013){Foreman-Mackey}, {Hogg}, {Lang}, \&
  {Goodman}}]{foremanmackey2013}
{Foreman-Mackey}, D., {Hogg}, D.~W., {Lang}, D., \& {Goodman}, J. 2013, \pasp,
  125, 306

\bibitem[{{Fragile} {et~al.}(2007){Fragile}, {Blaes}, {Anninos}, \&
  {Salmonson}}]{fragile2007}
{Fragile}, P.~C., {Blaes}, O.~M., {Anninos}, P., \& {Salmonson}, J.~D. 2007,
  \apj, 668, 417

\bibitem[{{F{\"u}rst} {et~al.}(2010){F{\"u}rst}, {Kreykenbohm}, {Pottschmidt},
  {Wilms}, {Hanke}, {Rothschild}, {Kretschmar}, {Schulz}, {Huenemoerder},
  {Klochkov}, \& {Staubert}}]{fuerst2010}
{F{\"u}rst}, F., {Kreykenbohm}, I., {Pottschmidt}, K., {et~al.} 2010, \aap,
  519, A37

\bibitem[{{Gandhi} {et~al.}(2016){Gandhi}, {Littlefair}, {Hardy}, {Dhillon},
  {Marsh}, {Shaw}, {Altamirano}, {Caballero-Garcia}, {Casares}, {Casella},
  {Castro-Tirado}, {Charles}, {Dallilar}, {Eikenberry}, {Fender}, {Hynes},
  {Knigge}, {Kuulkers}, {Mooley}, {Mu{\~n}oz-Darias}, {Pahari}, {Rahoui},
  {Russell}, {Hern{\'a}ndez Santisteban}, {Shahbaz}, {Terndrup}, {Tomsick}, \&
  {Walton}}]{gandhi2016}
{Gandhi}, P., {Littlefair}, S.~P., {Hardy}, L.~K., {et~al.} 2016, \mnras,
  arXiv:1603.04461

\bibitem[{{Garner} {et~al.}(2015){Garner}, {Eikenberry}, {Stelter}, {Raines},
  {Charcos}, {Edwards}, {Lasso-Cabrera}, {Marin-Franch}, {Cenarro}, {Bennett},
  {Mullin}, {Chinn}, {Ackley}, {Varosi}, {Warner}, {Frommeyer}, {Herlevich},
  {Miller}, {Murphey}, {Donoso}, {Vega}, {Packham}, {Dallilar}, {Scarpa},
  {Gerarts}, {Martin}, {Calero}, {Sanchez}, {Siegel}, {Losada}, {Perez},
  {Sendra}, \& {Acosta}}]{garner2015}
{Garner}, A., {Eikenberry}, S., {Stelter}, R.~D., {et~al.} 2015, The
  Astronomer's Telegram, 7663

\bibitem[{{Gazeas} {et~al.}(2015){Gazeas}, {Vasilopoulos}, {Petropoulou}, \&
  {Sapountzis}}]{gazeas2015}
{Gazeas}, K., {Vasilopoulos}, G., {Petropoulou}, M., \& {Sapountzis}, K. 2015,
  The Astronomer's Telegram, 7650

\bibitem[{{Gilfanov} {et~al.}(1999){Gilfanov}, {Churazov}, \&
  {Revnivtsev}}]{gilfanov1999}
{Gilfanov}, M., {Churazov}, E., \& {Revnivtsev}, M. 1999, \aap, 352, 182

\bibitem[{{Golenetskii} {et~al.}(2015){Golenetskii}, {Aptekar}, {Frederiks},
  {Pal'Shin}, {Oleynik}, {Ulanov}, {Svinkin}, {Tsvetkova}, {Lysenko}, \&
  {Cline}}]{golenetskii2015}
{Golenetskii}, S., {Aptekar}, R., {Frederiks}, D., {et~al.} 2015, GRB
  Coordinates Network, 17938

\bibitem[{{Grinberg} {et~al.}(2011){Grinberg}, {Kreykenbohm}, {F{\"u}rst},
  {Wilms}, {Pottschmidt}, {Cadolle Bel}, {Rodriguez}, {Marcu}, {Suchy},
  {Markowitz}, \& {Nowak}}]{grinberg2011}
{Grinberg}, V., {Kreykenbohm}, I., {F{\"u}rst}, F., {et~al.} 2011, Acta
  Polytechnica, 51, 33

\bibitem[{{Grinberg} {et~al.}(2014){Grinberg}, {Pottschmidt}, {B{\"o}ck},
  {Schmid}, {Nowak}, {Uttley}, {Tomsick}, {Rodriguez}, {Hell}, {Markowitz},
  {Bodaghee}, {Cadolle Bel}, {Rothschild}, \& {Wilms}}]{grinberg2014}
{Grinberg}, V., {Pottschmidt}, K., {B{\"o}ck}, M., {et~al.} 2014, \aap, 565, A1

\bibitem[{{Groth}(1975)}]{groth1975}
{Groth}, E.~J. 1975, \apjs, 29, 285

\bibitem[{Harrison {et~al.}(2013)Harrison, Craig, Christensen, Hailey, Zhang,
  Boggs, Stern, Cook, Forster, Giommi, Grefenstette, Kim, Kitaguchi, Koglin,
  Madsen, Mao, Miyasaka, Mori, Perri, Pivovaroff, Puccetti, Rana, Westergaard,
  Willis, Zoglauer, An, Bachetti, Barri{\`e}re, Bellm, Bhalerao, Brejnholt,
  Fuerst, Liebe, Markwardt, Nynka, Vogel, Walton, Wik, Alexander, Cominsky,
  Hornschemeier, Hornstrup, Kaspi, Madejski, Matt, Molendi, Smith, Tomsick,
  Ajello, Ballantyne, Balokovi{\'c}, Barret, Bauer, Blandford, Brandt,
  Brenneman, Chiang, Chakrabarty, Chenevez, Comastri, Dufour, Elvis, Fabian,
  Farrah, Fryer, Gotthelf, Grindlay, Helfand, Krivonos, Meier, Miller,
  Natalucci, Ogle, Ofek, Ptak, Reynolds, Rigby, Tagliaferri, Thorsett,
  Treister, \& Urry}]{nustar13}
Harrison, F.~A., Craig, W.~W., Christensen, F.~E., {et~al.} 2013, ApJ, 770, 103

\bibitem[{{Heil} {et~al.}(2015){Heil}, {Uttley}, \& {Klein-Wolt}}]{heil2015}
{Heil}, L.~M., {Uttley}, P., \& {Klein-Wolt}, M. 2015, \mnras, 448, 3348

\bibitem[{{Homan}(2012)}]{homan2012}
{Homan}, J. 2012, \apjl, 760, L30

\bibitem[{{Homan} {et~al.}(2001){Homan}, {Wijnands}, {van der Klis}, {Belloni},
  {van Paradijs}, {Klein-Wolt}, {Fender}, \& {M{\'e}ndez}}]{homan2001}
{Homan}, J., {Wijnands}, R., {van der Klis}, M., {et~al.} 2001, \apjs, 132, 377

\bibitem[{{Huppenkothen} {et~al.}(2013){Huppenkothen}, {Watts}, {Uttley}, {van
  der Horst}, {van der Klis}, {Kouveliotou}, {G{\"o}{\v g}{\"u}{\c s}},
  {Granot}, {Vaughan}, \& {Finger}}]{huppenkothen2013}
{Huppenkothen}, D., {Watts}, A.~L., {Uttley}, P., {et~al.} 2013, \apj, 768, 87

\bibitem[{{Inglis} {et~al.}(2015){Inglis}, {Ireland}, \&
  {Dominique}}]{inglis2015}
{Inglis}, A.~R., {Ireland}, J., \& {Dominique}, M. 2015, \apj, 798, 108

\bibitem[{{Ingram} \& {Done}(2011)}]{ingram2011}
{Ingram}, A., \& {Done}, C. 2011, \mnras, 415, 2323

\bibitem[{{Ingram} {et~al.}(2009){Ingram}, {Done}, \& {Fragile}}]{ingram2009}
{Ingram}, A., {Done}, C., \& {Fragile}, P.~C. 2009, \mnras, 397, L101

\bibitem[{{Ingram} \& {Motta}(2014)}]{ingram2014}
{Ingram}, A., \& {Motta}, S. 2014, \mnras, 444, 2065

\bibitem[{{Ingram} \& {van der Klis}(2013)}]{ingram2013}
{Ingram}, A., \& {van der Klis}, M. 2013, MNRAS, 434, 1476

\bibitem[{{Jenke} \& {Burns}(2015)}]{jenkegcn2015}
{Jenke}, P., \& {Burns}, E. 2015, GRB Coordinates Network, 17988

\bibitem[{{Jenke} {et~al.}(2016){Jenke}, {Wilson-Hodge}, {Homan}, {Veres},
  {Briggs}, {Burns}, {Connaughton}, {Finger}, \& {Hui}}]{jenke2016}
{Jenke}, P.~A., {Wilson-Hodge}, C.~A., {Homan}, J., {et~al.} 2016, ArXiv
  e-prints, arXiv:1601.00911

\bibitem[{{Kalamkar} {et~al.}(2015){Kalamkar}, {Casella}, {Uttley}, {O'Brien},
  {Russell}, {Maccarone}, {van der Klis}, \& {Vincentelli}}]{kalamkar2015}
{Kalamkar}, M., {Casella}, P., {Uttley}, P., {et~al.} 2015, ArXiv e-prints,
  arXiv:1510.08907

\bibitem[{{Kalamkar} {et~al.}(2013){Kalamkar}, {van der Klis}, {Uttley},
  {Altamirano}, \& {Wijnands}}]{kalamkar2013}
{Kalamkar}, M., {van der Klis}, M., {Uttley}, P., {Altamirano}, D., \&
  {Wijnands}, R. 2013, \apj, 766, 89

\bibitem[{{Kalemci} {et~al.}(2005){Kalemci}, {Tomsick}, {Buxton}, {Rothschild},
  {Pottschmidt}, {Corbel}, {Brocksopp}, \& {Kaaret}}]{kalemci2005}
{Kalemci}, E., {Tomsick}, J.~A., {Buxton}, M.~M., {et~al.} 2005, \apj, 622, 508

\bibitem[{{Kalemci} {et~al.}(2003){Kalemci}, {Tomsick}, {Rothschild},
  {Pottschmidt}, {Corbel}, {Wijnands}, {Miller}, \& {Kaaret}}]{kalemci2003}
{Kalemci}, E., {Tomsick}, J.~A., {Rothschild}, R.~E., {et~al.} 2003, \apj, 586,
  419

\bibitem[{{Kalemci} {et~al.}(2001){Kalemci}, {Tomsick}, {Rothschild},
  {Pottschmidt}, \& {Kaaret}}]{kalemci2001}
{Kalemci}, E., {Tomsick}, J.~A., {Rothschild}, R.~E., {Pottschmidt}, K., \&
  {Kaaret}, P. 2001, \apj, 563, 239

\bibitem[{{Khargharia} {et~al.}(2010){Khargharia}, {Froning}, \&
  {Robinson}}]{khargharia2010}
{Khargharia}, J., {Froning}, C.~S., \& {Robinson}, E.~L. 2010, \apj, 716, 1105

\bibitem[{{King} \& {Lasota}(2014)}]{king2014}
{King}, A., \& {Lasota}, J.-P. 2014, \mnras, 444, L30

\bibitem[{{King} {et~al.}(2015){King}, {Miller}, {Raymond}, {Reynolds}, \&
  {Morningstar}}]{king2015}
{King}, A.~L., {Miller}, J.~M., {Raymond}, J., {Reynolds}, M.~T., \&
  {Morningstar}, W. 2015, \apjl, 813, L37

\bibitem[{{Kitamoto} {et~al.}(1989){Kitamoto}, {Tsunemi}, {Miyamoto},
  {Yamashita}, \& {Mizobuchi}}]{kitamoto1989}
{Kitamoto}, S., {Tsunemi}, H., {Miyamoto}, S., {Yamashita}, K., \& {Mizobuchi},
  S. 1989, \nat, 342, 518

\bibitem[{{Klein-Wolt} \& {van der Klis}(2008)}]{kleinwolt2008}
{Klein-Wolt}, M., \& {van der Klis}, M. 2008, \apj, 675, 1407

\bibitem[{{Koljonen} {et~al.}(2011){Koljonen}, {Hannikainen}, \&
  {McCollough}}]{koljonen2011}
{Koljonen}, K.~I.~I., {Hannikainen}, D.~C., \& {McCollough}, M.~L. 2011,
  \mnras, 416, L84

\bibitem[{{K{\"o}rding} {et~al.}(2007){K{\"o}rding}, {Migliari}, {Fender},
  {Belloni}, {Knigge}, \& {McHardy}}]{koerding2007}
{K{\"o}rding}, E.~G., {Migliari}, S., {Fender}, R., {et~al.} 2007, \mnras, 380,
  301

\bibitem[{{Kotze} \& {Charles}(2012)}]{kotze2012}
{Kotze}, M.~M., \& {Charles}, P.~A. 2012, \mnras, 420, 1575

\bibitem[{{Kouveliotou} {et~al.}(1993){Kouveliotou}, {Finger}, {Fishman},
  {Meegan}, {Wilson}, {Paciesas}, {Minamitani}, \& {van
  Paradijs}}]{kouveliotou1993}
{Kouveliotou}, C., {Finger}, M.~H., {Fishman}, G.~J., {et~al.} 1993, in
  American Institute of Physics Conference Series, Vol. 280, American Institute
  of Physics Conference Series, ed. M.~{Friedlander}, N.~{Gehrels}, \& D.~J.
  {Macomb}, 319--323

\bibitem[{{Kuulkers} {et~al.}(2015){Kuulkers}, {Motta}, {Kajava}, {Homan},
  {Fender}, \& {Jonker}}]{kuulkers2015}
{Kuulkers}, E., {Motta}, S., {Kajava}, J., {et~al.} 2015, The Astronomer's
  Telegram, 7647

\bibitem[{{Lebrun} {et~al.}(2003){Lebrun}, {Leray}, {Lavocat}, {Cr{\'e}tolle},
  {Arqu{\`e}s}, {Blondel}, {Bonnin}, {Bou{\`e}re}, {Cara}, {Chaleil}, {Daly},
  {Desages}, {Dzitko}, {Horeau}, {Laurent}, {Limousin}, {Mathy}, {Mauguen},
  {Meignier}, {Molini{\'e}}, {Poindron}, {Rouger}, {Sauvageon}, \&
  {Tourrette}}]{lebrun2003}
{Lebrun}, F., {Leray}, J.~P., {Lavocat}, P., {et~al.} 2003, \aap, 411, L141

\bibitem[{{Liddle}(2007)}]{liddle2007}
{Liddle}, A.~R. 2007, \mnras, 377, L74

\bibitem[{{Lightman} \& {Eardley}(1974)}]{lightman1974}
{Lightman}, A.~P., \& {Eardley}, D.~M. 1974, \apjl, 187, L1

\bibitem[{{Lin} {et~al.}(2000){Lin}, {Smith}, {Liang}, \&
  {B{\"o}ttcher}}]{lin2000}
{Lin}, D., {Smith}, I.~A., {Liang}, E.~P., \& {B{\"o}ttcher}, M. 2000, \apjl,
  543, L141

\bibitem[{{Lubow} {et~al.}(2002){Lubow}, {Ogilvie}, \& {Pringle}}]{lubow2002}
{Lubow}, S.~H., {Ogilvie}, G.~I., \& {Pringle}, J.~E. 2002, \mnras, 337, 706

\bibitem[{{Lund} {et~al.}(2003){Lund}, {Budtz-J{\o}rgensen}, {Westergaard},
  {Brandt}, {Rasmussen}, {Hornstrup}, {Oxborrow}, {Chenevez}, {Jensen},
  {Laursen}, {Andersen}, {Mogensen}, {Rasmussen}, {Om{\o}}, {Pedersen},
  {Polny}, {Andersson}, {Andersson}, {K{\"a}m{\"a}r{\"a}inen}, {Vilhu},
  {Huovelin}, {Maisala}, {Morawski}, {Juchnikowski}, {Costa}, {Feroci},
  {Rubini}, {Rapisarda}, {Morelli}, {Carassiti}, {Frontera}, {Pelliciari},
  {Loffredo}, {Mart{\'{\i}}nez N{\'u}{\~n}ez}, {Reglero}, {Velasco}, {Larsson},
  {Svensson}, {Zdziarski}, {Castro-Tirado}, {Attina}, {Goria}, {Giulianelli},
  {Cordero}, {Rezazad}, {Schmidt}, {Carli}, {Gomez}, {Jensen}, {Sarri},
  {Tiemon}, {Orr}, {Much}, {Kretschmar}, \& {Schnopper}}]{lund2003}
{Lund}, N., {Budtz-J{\o}rgensen}, C., {Westergaard}, N.~J., {et~al.} 2003,
  \aap, 411, L231

\bibitem[{{Lyubarskii}(1997)}]{lyubarskii1997}
{Lyubarskii}, Y.~E. 1997, \mnras, 292, 679

\bibitem[{{Makino} {et~al.}(1989){Makino}, {Wagner}, {Starrfield}, {Buie},
  {Bond}, {Johnson}, {Harrison}, \& {Gehrz}}]{makino1989}
{Makino}, F., {Wagner}, R.~M., {Starrfield}, S., {et~al.} 1989, \iaucirc, 4786

\bibitem[{{Markoff} {et~al.}(2005){Markoff}, {Nowak}, \& {Wilms}}]{markoff2005}
{Markoff}, S., {Nowak}, M.~A., \& {Wilms}, J. 2005, \apj, 635, 1203

\bibitem[{{McHardy} {et~al.}(2006){McHardy}, {Koerding}, {Knigge}, {Uttley}, \&
  {Fender}}]{mchardy2006}
{McHardy}, I.~M., {Koerding}, E., {Knigge}, C., {Uttley}, P., \& {Fender},
  R.~P. 2006, \nat, 444, 730

\bibitem[{{Meegan} {et~al.}(2009){Meegan}, {Lichti}, {Bhat}, {Bissaldi},
  {Briggs}, {Connaughton}, {Diehl}, {Fishman}, {Greiner}, {Hoover}, {van der
  Horst}, {von Kienlin}, {Kippen}, {Kouveliotou}, {McBreen}, {Paciesas},
  {Preece}, {Steinle}, {Wallace}, {Wilson}, \& {Wilson-Hodge}}]{meegan2009}
{Meegan}, C., {Lichti}, G., {Bhat}, P.~N., {et~al.} 2009, \apj, 702, 791

\bibitem[{{Miller}(2007)}]{miller2007}
{Miller}, J.~M. 2007, \araa, 45, 441

\bibitem[{{Miller} {et~al.}(2001){Miller}, {Wijnands}, {Homan}, {Belloni},
  {Pooley}, {Corbel}, {Kouveliotou}, {van der Klis}, \& {Lewin}}]{miller2001}
{Miller}, J.~M., {Wijnands}, R., {Homan}, J., {et~al.} 2001, \apj, 563, 928

\bibitem[{{Miller} {et~al.}(2015){Miller}, {Tomsick}, {Bachetti}, {Wilkins},
  {Boggs}, {Christensen}, {Craig}, {Fabian}, {Grefenstette}, {Hailey},
  {Harrison}, {Kara}, {King}, {Stern}, \& {Zhang}}]{miller2015}
{Miller}, J.~M., {Tomsick}, J.~A., {Bachetti}, M., {et~al.} 2015, \apjl, 799,
  L6

\bibitem[{{Miller-Jones} {et~al.}(2009){Miller-Jones}, {Jonker}, {Dhawan},
  {Brisken}, {Rupen}, {Nelemans}, \& {Gallo}}]{millerjones2009}
{Miller-Jones}, J.~C.~A., {Jonker}, P.~G., {Dhawan}, V., {et~al.} 2009, \apjl,
  706, L230

\bibitem[{{Miyamoto} \& {Kitamoto}(1991)}]{miyamoto1991}
{Miyamoto}, S., \& {Kitamoto}, S. 1991, \apj, 374, 741

\bibitem[{{Mooley} {et~al.}(2015{\natexlab{a}}){Mooley}, {Clarke}, \&
  {Fender}}]{mooley2015b}
{Mooley}, K., {Clarke}, F., \& {Fender}, R. 2015{\natexlab{a}}, The
  Astronomer's Telegram, 7714

\bibitem[{{Mooley} {et~al.}(2015{\natexlab{b}}){Mooley}, {Fender}, {Anderson},
  {Staley}, {Kuulkers}, \& {Rumsey}}]{mooley2015a}
{Mooley}, K., {Fender}, R., {Anderson}, G., {et~al.} 2015{\natexlab{b}}, The
  Astronomer's Telegram, 7658

\bibitem[{{Motch} {et~al.}(1983){Motch}, {Ricketts}, {Page}, {Ilovaisky}, \&
  {Chevalier}}]{motch1983}
{Motch}, C., {Ricketts}, M.~J., {Page}, C.~G., {Ilovaisky}, S.~A., \&
  {Chevalier}, C. 1983, \aap, 119, 171

\bibitem[{{Motta} {et~al.}(2015{\natexlab{a}}){Motta}, {Beardmore}, {Oates},
  {Sanna}, {Kuulkers}, {Kajava}, \& {Sanchez-Fernanedz}}]{mottaatel2015}
{Motta}, S., {Beardmore}, A., {Oates}, S., {et~al.} 2015{\natexlab{a}}, The
  Astronomer's Telegram, 7665

\bibitem[{{Motta} {et~al.}(2012){Motta}, {Homan}, {Mu{\~n}oz Darias},
  {Casella}, {Belloni}, {Hiemstra}, \& {M{\'e}ndez}}]{motta2012}
{Motta}, S., {Homan}, J., {Mu{\~n}oz Darias}, T., {et~al.} 2012, \mnras, 427,
  595

\bibitem[{{Motta} {et~al.}(2015{\natexlab{b}}){Motta}, {Casella}, {Henze},
  {Mu{\~n}oz-Darias}, {Sanna}, {Fender}, \& {Belloni}}]{motta2015}
{Motta}, S.~E., {Casella}, P., {Henze}, M., {et~al.} 2015{\natexlab{b}},
  \mnras, 447, 2059

\bibitem[{{Mu{\~n}oz-Darias} {et~al.}(2011){Mu{\~n}oz-Darias}, {Motta}, \&
  {Belloni}}]{munosdarias2011}
{Mu{\~n}oz-Darias}, T., {Motta}, S., \& {Belloni}, T.~M. 2011, \mnras, 410, 679

\bibitem[{{Narayan} \& {Yi}(1995)}]{narayan1995}
{Narayan}, R., \& {Yi}, I. 1995, \apj, 452, 710

\bibitem[{{Natalucci} {et~al.}(2015){Natalucci}, {Fiocchi}, {Bazzano},
  {Ubertini}, {Roques}, \& {Jourdain}}]{natalucci2015}
{Natalucci}, L., {Fiocchi}, M., {Bazzano}, A., {et~al.} 2015, \apjl, 813, L21

\bibitem[{{Negoro} {et~al.}(2015){Negoro}, {Matsumitsu}, {Mihara}, {Serino},
  {Matsuoka}, {Nakahira}, {Ueno}, {Tomida}, {Kimura}, {Ishikawa}, {Nakagawa},
  {Sugizaki}, {Shidatsu}, {Sugimoto}, {Takagi}, {Kawai}, {Yoshii}, {Tachibana},
  {Yoshida}, {Sakamoto}, {Kawakubo}, {Ohtsuki}, {Tsunemi}, {Imatani},
  {Nakajima}, {Tanaka}, {Ueda}, {Kawamuro}, {Hori}, {Tsuboi}, {Kanetou},
  {Yamauchi}, {Itoh}, {Yamaoka}, \& {Morii}}]{negoro2015}
{Negoro}, H., {Matsumitsu}, T., {Mihara}, T., {et~al.} 2015, The Astronomer's
  Telegram, 7646

\bibitem[{{Neilsen} {et~al.}(2011){Neilsen}, {Remillard}, \&
  {Lee}}]{neilsen2011}
{Neilsen}, J., {Remillard}, R.~A., \& {Lee}, J.~C. 2011, \apj, 737, 69

\bibitem[{{Nowak}(2000)}]{nowak2000}
{Nowak}, M.~A. 2000, \mnras, 318, 361

\bibitem[{{Nowak} {et~al.}(1999){Nowak}, {Wilms}, \& {Dove}}]{nowak1999}
{Nowak}, M.~A., {Wilms}, J., \& {Dove}, J.~B. 1999, \apj, 517, 355

\bibitem[{{Nowak} {et~al.}(2002){Nowak}, {Wilms}, \& {Dove}}]{nowak2002}
---. 2002, \mnras, 332, 856

\bibitem[{{Oosterbroek} {et~al.}(1996){Oosterbroek}, {van der Klis}, {Vaughan},
  {van Paradijs}, {Rutledge}, {Lewin}, {Tanaka}, {Nagase}, {Dotani}, {Mitsuda},
  \& {Yoshida}}]{Oosterbroek1996}
{Oosterbroek}, T., {van der Klis}, M., {Vaughan}, B., {et~al.} 1996, \aap, 309,
  781

\bibitem[{{Oosterbroek} {et~al.}(1997){Oosterbroek}, {van der Klis}, {van
  Paradijs}, {Vaughan}, {Rutledge}, {Lewin}, {Tanaka}, {Nagase}, {Dotani},
  {Mitsuda}, \& {Miyamoto}}]{oosterbroek1997}
{Oosterbroek}, T., {van der Klis}, M., {van Paradijs}, J., {et~al.} 1997, \aap,
  321, 776

\bibitem[{{Parker} {et~al.}(2015){Parker}, {Tomsick}, {Miller}, {Yamaoka},
  {Lohfink}, {Nowak}, {Fabian}, {Alston}, {Boggs}, {Christensen}, {Craig},
  {F{\"u}rst}, {Gandhi}, {Grefenstette}, {Grinberg}, {Hailey}, {Harrison},
  {Kara}, {King}, {Stern}, {Walton}, {Wilms}, \& {Zhang}}]{parker2015}
{Parker}, M.~L., {Tomsick}, J.~A., {Miller}, J.~M., {et~al.} 2015, \apj, 808, 9

\bibitem[{{Pasham} {et~al.}(2013){Pasham}, {Strohmayer}, \&
  {Mushotzky}}]{pasham2013}
{Pasham}, D.~R., {Strohmayer}, T.~E., \& {Mushotzky}, R.~F. 2013, \apjl, 771,
  L44

\bibitem[{{Pottschmidt} {et~al.}(2003){Pottschmidt}, {Wilms}, {Nowak},
  {Pooley}, {Gleissner}, {Heindl}, {Smith}, {Remillard}, \&
  {Staubert}}]{pottschmidt2003}
{Pottschmidt}, K., {Wilms}, J., {Nowak}, M.~A., {et~al.} 2003, \aap, 407, 1039

\bibitem[{{Pringle}(1996)}]{pringle1996}
{Pringle}, J.~E. 1996, \mnras, 281, 357

\bibitem[{{Prosvetov} \& {Grebenev}(2015)}]{prosvetov2015}
{Prosvetov}, A.~V., \& {Grebenev}, S.~A. 2015, The Astronomer's Telegram, 7726

\bibitem[{{Radhika} {et~al.}(2016){Radhika}, {Nandi}, {Agrawal}, \&
  {Mandal}}]{radhika2016}
{Radhika}, D., {Nandi}, A., {Agrawal}, V.~K., \& {Mandal}, S. 2016, ArXiv
  e-prints, arXiv:1601.03234

\bibitem[{{Raman} {et~al.}(2016){Raman}, {Paul}, {Bhattacharya}, \&
  {Mohan}}]{raman2016}
{Raman}, G., {Paul}, B., {Bhattacharya}, D., \& {Mohan}, V. 2016, \mnras, 458,
  1302

\bibitem[{{Remillard} {et~al.}(1999{\natexlab{a}}){Remillard}, {McClintock},
  {Sobczak}, {Bailyn}, {Orosz}, {Morgan}, \& {Levine}}]{remillard1999a}
{Remillard}, R.~A., {McClintock}, J.~E., {Sobczak}, G.~J., {et~al.}
  1999{\natexlab{a}}, \apjl, 517, L127

\bibitem[{{Remillard} {et~al.}(1999{\natexlab{b}}){Remillard}, {Morgan},
  {McClintock}, {Bailyn}, \& {Orosz}}]{remillard1999b}
{Remillard}, R.~A., {Morgan}, E.~H., {McClintock}, J.~E., {Bailyn}, C.~D., \&
  {Orosz}, J.~A. 1999{\natexlab{b}}, \apj, 522, 397

\bibitem[{{Remillard} {et~al.}(2002){Remillard}, {Sobczak}, {Muno}, \&
  {McClintock}}]{remillard2002}
{Remillard}, R.~A., {Sobczak}, G.~J., {Muno}, M.~P., \& {McClintock}, J.~E.
  2002, \apj, 564, 962

\bibitem[{{Revnivtsev} {et~al.}(2000){Revnivtsev}, {Gilfanov}, \&
  {Churazov}}]{revnivtsev2000}
{Revnivtsev}, M., {Gilfanov}, M., \& {Churazov}, E. 2000, \aap, 363, 1013

\bibitem[{{Richter}(1989)}]{richter1989}
{Richter}, G.~A. 1989, Information Bulletin on Variable Stars, 3362

\bibitem[{{Rodriguez} {et~al.}(2015){Rodriguez}, {Cadolle Bel},
  {Alfonso-Garz{\'o}n}, {Siegert}, {Zhang}, {Grinberg}, {Savchenko}, {Tomsick},
  {Chenevez}, {Clavel}, {Corbel}, {Diehl}, {Domingo}, {Gouiff{\`e}s},
  {Greiner}, {Krause}, {Laurent}, {Loh}, {Markoff}, {Mas-Hesse},
  {Miller-Jones}, {Russell}, \& {Wilms}}]{rodriguez2015}
{Rodriguez}, J., {Cadolle Bel}, M., {Alfonso-Garz{\'o}n}, J., {et~al.} 2015,
  \aap, 581, L9

\bibitem[{{Roques} {et~al.}(2015){Roques}, {Jourdain}, {Bazzano}, {Fiocchi},
  {Natalucci}, \& {Ubertini}}]{roques2015}
{Roques}, J.-P., {Jourdain}, E., {Bazzano}, A., {et~al.} 2015, \apjl, 813, L22

\bibitem[{{Schnittman} {et~al.}(2006){Schnittman}, {Homan}, \&
  {Miller}}]{schnittman2006}
{Schnittman}, J.~D., {Homan}, J., \& {Miller}, J.~M. 2006, \apj, 642, 420

\bibitem[{Schwarz {et~al.}(1978)}]{schwarz1978}
Schwarz, G., {et~al.} 1978, The annals of statistics, 6, 461

\bibitem[{{Shahbaz} {et~al.}(1994){Shahbaz}, {Ringwald}, {Bunn}, {Naylor},
  {Charles}, \& {Casares}}]{shahbaz1994}
{Shahbaz}, T., {Ringwald}, F.~A., {Bunn}, J.~C., {et~al.} 1994, \mnras, 271,
  L10

\bibitem[{{Sivakoff} {et~al.}(2015){Sivakoff}, {Bahramian}, {Altamirano},
  {Beardmore}, {Kuulkers}, \& {Motta}}]{sivakoffatel2015}
{Sivakoff}, G.~R., {Bahramian}, A., {Altamirano}, D., {et~al.} 2015, The
  Astronomer's Telegram, 7959

\bibitem[{{Smith} {et~al.}(1997){Smith}, {Heindl}, {Swank}, {Leventhal},
  {Mirabel}, \& {Rodriguez}}]{smith1997}
{Smith}, D.~M., {Heindl}, W.~A., {Swank}, J., {et~al.} 1997, \apjl, 489, L51

\bibitem[{{Sobczak} {et~al.}(2000){Sobczak}, {Remillard}, {Muno}, \&
  {McClintock}}]{sobczak2000}
{Sobczak}, G.~J., {Remillard}, R.~A., {Muno}, M.~P., \& {McClintock}, J.~E.
  2000, ArXiv Astrophysics e-prints, astro-ph/0004215

\bibitem[{{Stella} \& {Vietri}(1998)}]{stella1998}
{Stella}, L., \& {Vietri}, M. 1998, \apjl, 492, L59

\bibitem[{{Strohmayer}(2001)}]{strohmayer2001}
{Strohmayer}, T.~E. 2001, \apjl, 552, L49

\bibitem[{{Takizawa} {et~al.}(1997){Takizawa}, {Dotani}, {Mitsuda}, {Matsuba},
  {Ogawa}, {Aoki}, {Asai}, {Ebisawa}, {Makishima}, {Miyamoto}, {Iga},
  {Vaughan}, {Rutledge}, \& {Lewin}}]{takizawa1997}
{Takizawa}, M., {Dotani}, T., {Mitsuda}, K., {et~al.} 1997, \apj, 489, 272

\bibitem[{{Tetarenko} {et~al.}(2015){Tetarenko}, {Sivakoff}, {Gurwell},
  {Petitpas}, {Wouterloot}, \& {Miller-Jones}}]{tetarenko2015}
{Tetarenko}, A., {Sivakoff}, G.~R., {Gurwell}, M.~A., {et~al.} 2015, The
  Astronomer's Telegram, 7661

\bibitem[{{Tomsick} {et~al.}(2009){Tomsick}, {Yamaoka}, {Corbel}, {Kaaret},
  {Kalemci}, \& {Migliari}}]{tomsick2009}
{Tomsick}, J.~A., {Yamaoka}, K., {Corbel}, S., {et~al.} 2009, \apjl, 707, L87

\bibitem[{{Tremaine} \& {Davis}(2014)}]{tremain2014}
{Tremaine}, S., \& {Davis}, S.~W. 2014, \mnras, 441, 1408

\bibitem[{{Ubertini} {et~al.}(2003){Ubertini}, {Lebrun}, {Di Cocco}, {Bazzano},
  {Bird}, {Broenstad}, {Goldwurm}, {La Rosa}, {Labanti}, {Laurent}, {Mirabel},
  {Quadrini}, i~{Ramsey}, {Reglero}, {Sabau}, {Sacco}, {Staubert}, {Vigroux},
  {Weisskopf}, \& {Zdziarski}}]{ubertini2003}
{Ubertini}, P., {Lebrun}, F., {Di Cocco}, G., {et~al.} 2003, \aap, 411, L131

\bibitem[{{Uttley} \& {Casella}(2014)}]{uttley2014}
{Uttley}, P., \& {Casella}, P. 2014, \ssr, 183, 453

\bibitem[{{Uttley} {et~al.}(2005){Uttley}, {McHardy}, \&
  {Vaughan}}]{uttley2005}
{Uttley}, P., {McHardy}, I.~M., \& {Vaughan}, S. 2005, \mnras, 359, 345

\bibitem[{{van der Klis}(1989)}]{vanderklis1989}
{van der Klis}, M. 1989, in NATO Advanced Science Institutes (ASI) Series C,
  Vol. 262, NATO Advanced Science Institutes (ASI) Series C, ed.
  H.~{{\"O}gelman} \& E.~P.~J. {van den Heuvel}, 27

\bibitem[{{van der Klis}(1997)}]{vanderklis1997}
{van der Klis}, M. 1997, in Statistical Challenges in Modern Astronomy II, ed.
  G.~J. {Babu} \& E.~D. {Feigelson}, 321

\bibitem[{{van Kerkwijk} {et~al.}(1992){van Kerkwijk}, {Charles}, {Geballe},
  {King}, {Miley}, {Molnar}, {van den Heuvel}, {van der Klis}, \& {van
  Paradijs}}]{vankerkwijk1992}
{van Kerkwijk}, M.~H., {Charles}, P.~A., {Geballe}, T.~R., {et~al.} 1992, \nat,
  355, 703

\bibitem[{{Vaughan}(2010)}]{vaughan2010}
{Vaughan}, S. 2010, \mnras, 402, 307

\bibitem[{{Vikhlinin} {et~al.}(1994){Vikhlinin}, {Churazov}, {Gilfanov},
  {Sunyaev}, {Dyachkov}, {Khavenson}, {Kremnev}, {Sukhanov}, {Ballet},
  {Laurent}, {Salotti}, {Claret}, {Olive}, {Denis}, {Mandrou}, \&
  {Roques}}]{vikhlinin1994}
{Vikhlinin}, A., {Churazov}, E., {Gilfanov}, M., {et~al.} 1994, \apj, 424, 395

\bibitem[{{Wachmann}(1948)}]{wachmann1948}
{Wachmann}, A.~A. 1948, {Beobachtungen von Veranderlichen in der Umgebung von
  Kapteyn-Feldern der nordlichen Milchstrasse.}

\bibitem[{{Walton} {et~al.}(2016){Walton}, {Mooley}, {King}, {Tomsick},
  {Miller}, {Dauser}, {Garcia}, {Bachetti}, {Brightman}, {Fabian}, {Forster},
  {Fuerst}, {Gandhi}, {Grefenstette}, {Harrison}, {Madsen}, {Meier},
  {Middleton}, {Natalucci}, {Rahoui}, {Rana}, \& {Stern}}]{walton2016}
{Walton}, D.~J., {Mooley}, K., {King}, A.~L., {et~al.} 2016, ArXiv e-prints,
  arXiv:1609.01293

\bibitem[{{Wijnands} {et~al.}(1999){Wijnands}, {Homan}, \& {van der
  Klis}}]{wijnands1999}
{Wijnands}, R., {Homan}, J., \& {van der Klis}, M. 1999, \apjl, 526, L33

\bibitem[{{Winkler} {et~al.}(2003){Winkler}, {Courvoisier}, {Di Cocco},
  {Gehrels}, {Gim{\'e}nez}, {Grebenev}, {Hermsen}, {Mas-Hesse}, {Lebrun},
  {Lund}, {Palumbo}, {Paul}, {Roques}, {Schnopper}, {Sch{\"o}nfelder},
  {Sunyaev}, {Teegarden}, {Ubertini}, {Vedrenne}, \& {Dean}}]{winkler2003}
{Winkler}, C., {Courvoisier}, T.~J.-L., {Di Cocco}, G., {et~al.} 2003, \aap,
  411, L1

\bibitem[{{Younes} {et~al.}(2015){Younes}, {Kouveliotou}, {Grefenstette},
  {Tomsick}, {Tennant}, {Finger}, {F{\"u}rst}, {Pottschmidt}, {Bhalerao},
  {Boggs}, {Boirin}, {Chakrabarty}, {Christensen}, {Craig}, {Degenaar},
  {Fabian}, {Gandhi}, {G{\"o}{\u g}{\"u}{\c s}}, {Hailey}, {Harrison},
  {Kennea}, {Miller}, {Stern}, \& {Zhang}}]{younes2015}
{Younes}, G., {Kouveliotou}, C., {Grefenstette}, B.~W., {et~al.} 2015, \apj,
  804, 43

\bibitem[{{{\.Z}ycki} {et~al.}(1999){{\.Z}ycki}, {Done}, \&
  {Smith}}]{zycki1999}
{{\.Z}ycki}, P.~T., {Done}, C., \& {Smith}, D.~A. 1999, \mnras, 309, 561

\end{thebibliography}
\bibliographystyle{apj}

\end{document}